\documentclass[journal]{IEEEtran}
\usepackage{multirow,diagbox}
\usepackage{verbatim}
\usepackage{graphicx}
\usepackage{color}
\usepackage{colortbl}
\usepackage{makecell}
\usepackage{cellspace}
\usepackage{stfloats}
\usepackage{longtable}

\usepackage{array,multirow,graphicx}
\usepackage{color}
\usepackage{cite}
\usepackage{float}
\usepackage{makecell}
\usepackage[x11names,dvipsnames,table]{xcolor} 
\usepackage{colortbl}
\usepackage{multirow}
\usepackage{subcaption}
\usepackage{lipsum}
\definecolor{mygray}{gray}{0.6}
\definecolor{myblue}{rgb}{0.8,0.85,1}
\usepackage{array}
\newcolumntype{L}[1]{>{\raggedright\let\newline\\\arraybackslash\hspace{0pt}}m{#1}}
\newcolumntype{C}[1]{>{\centering\let\newline\\\arraybackslash\hspace{0pt}}m{#1}}
\newcolumntype{R}[1]{>{\raggedleft\let\newline\\\arraybackslash\hspace{0pt}}m{#1}}
\newcommand{\noun}[1]{\textsc{#1}}
\usepackage{array, tabularx, boldline}
\usepackage{eurosym}
\usepackage{amstext} 
\DeclareRobustCommand{\officialeuro}{%
  \ifmmode\expandafter\text\fi
  {\fontencoding{U}\fontfamily{eurosym}\selectfont e}}

\usepackage{array, tabularx, boldline}
\usepackage{graphicx}
\usepackage{cellspace}

\usepackage{cite}
\usepackage{url}

\ifCLASSINFOpdf
\else
\fi
\usepackage{amsmath}
\usepackage{bm}
\usepackage{mathtools}          
\usepackage{amssymb}

\usepackage{array}
\makeatletter
\def\ps@IEEEtitlepagestyle{
  \def\@oddfoot{\mycopyrightnotice}
  \def\@evenfoot{}
}
\def\mycopyrightnotice{
  {\footnotesize
  \begin{minipage}{\textwidth}
  \centering
  Copyright~\copyright~2019 IEEE. Personal use of this material is permitted. Permission from IEEE must be obtained for all other uses, in any current or future media, including reprinting/republishing this material for advertising or promotional purposes, creating new collective works, for resale or redistribution to servers or lists, or reuse of any copyrighted component of this work in other works.
  \end{minipage}
  }
}

\begin{document}
%
\title{A Survey on Applications of Game Theory in Blockchain}
%
%
%


\author{
  \IEEEauthorblockN{Ziyao Liu,
  Nguyen Cong Luong,
  Wenbo Wang,~\IEEEmembership{Member,~IEEE,}
  Dusit Niyato,~\IEEEmembership{Fellow,~IEEE,}
  Ping Wang,~\IEEEmembership{Senior Member,~IEEE,}
  Ying-Chang Liang,~\IEEEmembership{Fellow,~IEEE,}
  and
  Dong In Kim,~\IEEEmembership{Fellow,~IEEE}
  \vspace*{-4mm}}
\vspace*{-4mm}

\thanks{Ziyao Liu, Nguyen Cong Luong, Wenbo Wang, and Dusit Niyato are with the School of Computer Science and Engineering, Nanyang Technological University, Singapore 639798
(email: ziyao002@ntu.edu.sg, clnguyen@ntu.edu.sg, wbwang@ntu.edu.sg, dniyato@ntu.edu.sg).}
\thanks{Ping Wang is with the Department of Electrical Engineering \& Computer Science, Lassonde School of Engineering, York University, 4700 Keele St., LAS 2016
Toronto, ON M3J 1P3, Canada (email: pingw@yorku.ca).}
\thanks{Y.-C. Liang is with Center for Intelligent Networking and Communications (CINC), University of Electronic Science and Technology of China, Chengdu, China. (email: liangyc@ieee.org).}
\thanks{D. I. Kim is with Department of Electrical and Computer Engineering, Sungkyunkwan University, 2066 Seobu-ro, Jangan-gu, Suwon 16419, Korea (email: dikim@skku.ac.kr).}

}
\maketitle



\begin{abstract}

In the past decades, the blockchain technology has attracted tremendous attention from both academia and industry. The popularity of blockchain networks was originated from a crypto-currency to serve as a decentralized and tamperproof transaction data ledger. Nowadays, blockchain as the key framework in the decentralized public data-ledger, has been applied to a wide range of scenarios far beyond crypto-currencies, such as Internet of Things (IoT), healthcare, and insurance. This survey aims to fill the gap between the large number of studies on blockchain network, where game theory emerges as an analytical tool, and the lack of a comprehensive survey on the game theoretical approaches applied in blockchain related issues. In this paper, we review game models proposed to address common issues in the blockchain network. The issues include security issues, e.g., selfish mining, majority attack and Denial of Service (DoS) attack, issues regard mining management, e.g., computational power allocation, reward allocation, and pool selection, as well as issues regarding blockchain economic and energy trading. Additionally, we discuss advantages and disadvantages of these selected game models and solutions. Finally, we highlight important challenges and future research directions of applying game theoretical approaches to incentive mechanism design, and the combination of blockchain with other technologies.

\end{abstract}

%
\IEEEpeerreviewmaketitle

\section{Introduction}

In the past decade, with the popularity of digital crypto-currencies, e.g., Bitcoin \cite{nakamoto2008bitcoin}, blockchain technology has attracted tremendous attention from both academia and industry \cite{zheng2016blockchain}. The blockchain was first proposed in \cite{nakamoto2008bitcoin} to serve as a crypto-currency transaction ledger, and is currently widely adopted for a large number of crypto-currencies, such as Ethereum \cite{wood2014ethereum}, Ripple \cite{schwartz2014ripple}, and EOS \cite{cox2017eos}. The blockchain technology guarantees the tamperproof ledger, transparent transactions, and trustless but secure tradings in a decentralized network. Thus, the blockchain network is recently applied in a wide range of scenarios far beyond crypto-currencies, such as Internet of Things (IoT) \cite{Tangle2018}, healthcare \cite{yue2016healthcare}, and insurance \cite{raikwar2018blockchain}. In general, blockchain is a distributed public data-ledger maintained by achieving the consensus among a number of nodes in a Peer-to-Peer (P2P) network. More specifically, the verified transaction data is stored in a chain of blocks, i.e., a basic data structure of blockchain, and the chain grows in an append-only manner with all new verified blocks to it. This process involves several operations such as verifying transactions, disseminating blocks, and attaching blocks to the blockchain.

As such, the blockchain requires a number of consensus nodes to participate in the network. The rational nodes perform actions or strategies that aim to maximize their own utility. Moreover, the malicious nodes may launch attacks that damage the blockchain networks. To address these security challenges, consensus protocols such as Byzantine Fault Tolerance (BFT) protocol \cite{castro2002practical} can be adopted. However, the consensus protocols require a centralized permission controller and only achieve the consensus among a very small group of nodes. Such a consensus protocol is thus not applicable to the blockchain network that is a decentralized and large-scale system. Different optimization approaches and solutions, e.g., a Markov Decision Process (MDP) \cite{altman1999constrained}, are used to analyze and optimize strategies of the blockchain nodes to prevent their misbehaviors. However, the optimization approaches do not take into account the interactions among the nodes. Recently, game theory \cite{myerson2013game} has been applied as an alternative solution in the blockchain network. Game theory is a study of mathematical models of strategic interaction between rational decision-makers \cite{han2012game}. Thus, game theory can be used to analyze the strategies of the consensus nodes as well as the interactions among them. Through the game theoretical analysis, the nodes can learn and predict mining behaviors\footnote{In blockchain systems which incentive nodes to participate in the consensus process of data record with digital tokens, the consensus nodes are frequently referred as block miners and their operations are referred as mining.} of each other, then having optimal reaction strategies based on equilibrium analysis. Moreover, game theory can be utilized to develop incentive mechanisms that discourage the nodes from executing misbehaviors or launching attacks. As such, game theory is natural in the decision making of all the consensus nodes in the blockchain networks.

Currently, there are some surveys related to the blockchain. However, the existing surveys do not discuss the applications of the game theory in the blockchain. In particular, the survey in \cite{tschorsch2016bitcoin} provides a comprehensive introduction of bitcoin network, the surveys in \cite{conti2018survey}, \cite{khalilov2018survey}, \cite{salman2018security} present security and privacy issues in the bitcoin network, the survey in \cite{ali2018applications} presents the blockchain applications on Internet of Things (IoT), the survey in \cite{yang2019integrated} discusses the integrations of blockchain and edge computing. To the best of our knowledge, there is no survey specifically discussing the use of game theory, as an efficient analysis tool, in blockchain networks. This motivates us to deliver the survey with the comprehensive literature review on the game models in the blockchain network. For convenience, the related works in this survey are classified based on issues in the blockchain network. The major issues consist of
(i) security issues such as selfish mining attacks and Denial-of-Service (DoS) attacks, (ii) mining management issues such as computational power allocation, fork chain selection, pool selection, and reward allocation, and (iii) applications atop the blockchain such as energy trading.

The rest of this paper is organized as follows. Section II briefly describes the general architecture of blockchain. Section III presents the fundamentals of game theory and game models that are commonly used in blockchain. Section IV discusses applications of game theory for security issues in blockchain. Section V presents applications of game theory for the mining management in blockchain. Section VI discusses applications of game theory atop blockchain platforms. Section VII outlines challenges and future research directions. Section VIII summarizes and concludes the paper.

\section{Overview and Fundamentals of Blockchain}\label{blkoverview}

In this section, we give an overview of blockchain on its concepts, data organization, working mechamism, and incentive compatibility.

\subsection{Overview of Blockchain}

The blockchain was first proposed as a decentralized tamperproof ledger which records a set of transactions. These transactions are verified through a decentralized consensus process among the trustless agents before attaching to the chain. Here, we summarize the key advantages that blockchain networks can offer as follows.

\begin{itemize}
\item Decentralized network: Due to the distributed network which allows every computing unit to utilize its computational power to take part in the blockchain, and that each transaction in the blockchain must achieve the agreement among all the nodes through the consensus protocol, the monopoly in centralized network can be removed in the blockchain.
\item Tamperproof ledger: The cryptographic techniques  used in blockchain ensure that any change on the transaction data in blockchain can be observed by all the nodes in the network. This means that the transaction recorded in the blockchain cannot be altered and tampered, unless the majority of nodes are compromised.
\item Transparent transaction: All the transactions in the blockchain can be traced back for verification, and these transactions are transparent to all the nodes in the blockchain network.
\item Trustless but secure trading: By using the digital asymmetric key signature, the blockchain network guarantees that only the sender and receiver which possess the pair of asymmetric key can execute the transaction, without intervention of any trust third-party.
\end{itemize}

\subsection{Data Organization and Workflow of Blockchain}\label{property}

Cryptographic data organization plays an extremely important role in the blockchain structure. We first introduce some basic components supporting the data organization within blockchain networks.

\begin{itemize}
\item Transaction: Transaction is the most basic component of blockchain. A transaction is proposed by the blockchain user and is composed of the transaction data which specifies the value in concern, e.g., the digital tokens in a crypto-currency, the addresses of the sender and the receiver, as well as the corresponding transaction fee \cite{nakamoto2008bitcoin}.
\item Block: A block is composed of a block header and a certain amount of transactions. The block header specifies the hash pointer and merkle tree data structure.
\item Hash pointer \cite{tschorsch2016bitcoin}: The hash pointer of the current block contains the hash value associated with the previous block, which also contains the hash pointer to the block before that one. Thereby, the hash pointers can be used to build a link of records, i.e., blockchain.
\item Merkle Tree \cite{merkle1987digital}: A merkle tree or hash tree is a tree in which each leaf node is marked by the hash value of the transaction data of a block, and those non-leaf nodes are marked by the hash value of the concatenation of its child nodes. This structure makes it impossible to tamper the data in blockchain privately.
\end{itemize}

\begin{figure}[htbp]
 \centering
\includegraphics[width=8 cm, height=4cm]{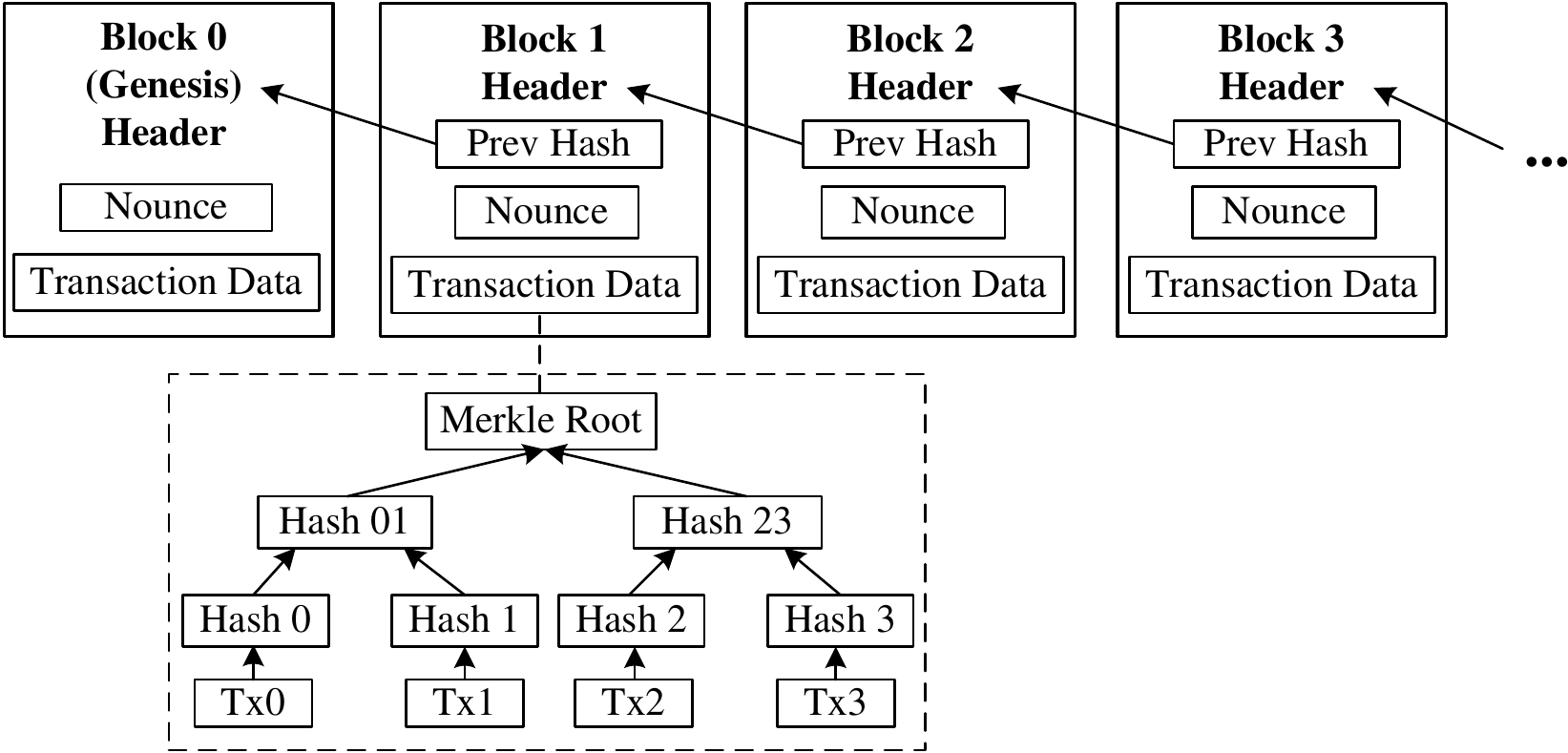}
 \caption{ An illustrative example of blockchain data structure where the transactions are included in the block and the block is represented by a merkle root.}
 \label{blockchaindatastructure}
\end{figure}

As shown in Fig. \ref{blockchaindatastructure}, a typical blockchain is an appending-only, ever-growing list of blocks, which are linked sequentially using the hash pointers as a linear linked list. More specifically, the block header includes a hash pointer which is associated with the previous block, and transaction data is represented as merkle trees.

\begin{figure}[htbp]
 \centering
\includegraphics[width=8.5 cm, height=2.5cm]{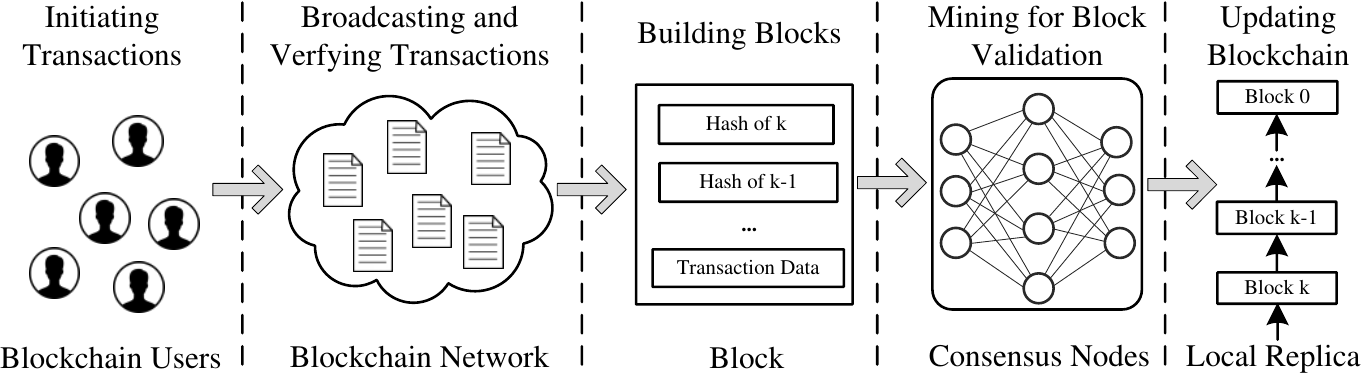}
 \caption{An overview of the blockchain workflow.}
 \label{blockchainlayer}
\end{figure}

Atop the basic cryptographic data organization, maintaining the blockchain network needs blockchain nodes to disseminate the transaction, store the data into blocks, verify the transaction, and eventually reach a consensus. The blockchain working mechanism works as follows (see Fig. \ref{blockchainlayer}).

\begin{itemize}

\item An initiated transaction is broadcast to the distributed network through a node.
\item The nodes in the blockchain verify the transaction as well as the node which broadcasts the transaction.
\item More than one node may bundle different subset of newly verified transactions into their candidate blocks and broadcast them the to the entire network.
\item All or part of the nodes in the blockchain network participate in the block validation by executing some certain functions defined by the consensus protocol.
\item The verified block is attached to the blockchain, and every node updates its local replica, i.e., the local views of whole ledger-data, of the blockchain.

\end{itemize}

In general, not all the nodes can be authenticated to join the blockchain network to execute the consensus protocol. According to the access control scheme \cite{buterin2015public} that determines which node can join the network, the blockchain platforms are classified into permissionless schemes, i.e., public blockchains, and permissioned schemes including private and consortium blockchains. When choosing the permissioned access control scheme, e.g., Hyperledger fabric \cite{cachin2016architecture}, the consensus needs to be reached among only a small group of authenticated nodes, and thus the permissioned blockchain network usually adopts BFT protocols, e.g., Byzantine Paxos \cite{cachin2009yet}. On the contrary, in permissionless blockchain, e.g., Ethereum \cite{wood2014ethereum}, any node can participate in the network, and some other consensus protocols are applied, such as Proof of Work (PoW) and Proof of Stake (PoS). Here we list some widely-used blockchain platforms and their consensus protocols in Table \ref{platform}.

\begin{table}[htbp]
\caption{Some Widely-used Blockchain Platforms}\label{platform}
\centering
 \begin{tabular}{| l | l | l |}
 \hline
 \textbf{Platform Name} & \textbf{Ledger type} & \textbf{Consensus Protocol} \\
 \hline
 Bitcoin \cite{nakamoto2008bitcoin} & Public & Proof of Work (PoW) \\
 \hline
 Ethereum \cite{wood2014ethereum} & Public & PoW \& Proof of Stake \\
 \hline
 Hyperledger Fabric \cite{cachin2016architecture} & Consortium & Pluggable algorithm \\
 \hline
 EOS \cite{cox2017eos} & Private & Delegated Proof of Stake \\
 \hline
 Stellar \cite{mazieres2015stellar} & Public \& Private & Stellar consensus protocol \\
 \hline
 Quorum \cite{moganquorum} & Private & Majority voting \\
 \hline
 Ripple \cite{schwartz2014ripple} & Private & Probabilistic voting \\
 \hline
 \end{tabular}
\end{table}

\subsection{Incentive Compatibility within Blockchain}

In blockchain network, the consensus protocol guarantees achieving the agreement among the nodes. A reliable consensus protocol needs to satisfy properties \cite{wang2018survey}. (i) Correctness: each node adopts the content and the order of transactions in the confirmed blockchain structure. (ii) Consistency: each node updates its local blockchain structure if a new block header is confirmed. (iii) Traceability: all transactions can be traced back for confirmation. However, in some case, disagreements may exist among the nodes. For example, the local blockchain replica of all the nodes are unable to be synchronized simultaneously due to the distributed network. Under this case, the nodes might maintain different blockchain ledgers, and thereby the fork chains appear. This means that the nodes may deviate from the protocol of maintaining the longest chain\footnote{ Due to the different strategies that nodes make to maximize their own utilities, the nodes may attach new verified blocks to the different blocks in blockchain, and thereby fork chains appear. The consensus protocols regulate the nodes to apply their work on the longest chain.}. Therefore, the blockchain consensus protocol is expected to be \textit{incentive compatible} \cite{wang2018survey}. This means that any node will suffer from financial loss, e.g., waste of investment in mining machine, whenever the node deviates from the protocol.

Currently, the most popular blockchain consensus protocol is the PoW-based Nakamoto consensus protocol \cite{nakamoto2008bitcoin}. The Nakamoto protocol achieves the consensus by solving a mathematical puzzle, i.e., by finding a hash value which satisfies a certain condition. The first node that solves the puzzle can broadcast the verified block to the blockchain network, and obtains the reward and the transaction fee. This process of solving puzzle and obtaining the reward is called mining. The design of the mining mechanism relies on both cryptography \cite{katz1996handbook} and game theory \cite{han2012game}.

Although the PoW protocol is widely used among the blockchain platforms, the incentive compatibility of the protocol has been openly questioned from game theoretical perspectives \cite{li2017survey}. The reason is that achieving the Nakamoto consensus involves nodes joining the network, executing the protocol, and maintaining the ledger. The nodes may deviate from the protocol to increase their own utilities. For example, the node may not broadcast its newly discovered blocks but choose to withhold the block to increase its utility \cite{wang2018survey}. The node trades off between the cost of withholding the block which is associated with the other nodes' strategies, and the mining reward and then chooses its strategy. To analyze the interactions among these consensus nodes, the game theoretical models (see Section \ref{background}) are developed and applied \cite{kroll2013economics}. In addition to the security issues, nodes' mining management in blockchain, e.g., computational power allocation \cite{dimitri2017bitcoin} and reward allocation \cite{schrijvers2016incentive}, adopt game models for the analysis as well. Apart from the Nakamoto protocol, game models are also widely used for analyzing the incentive compatibility with other consensus protocols, e.g., Proof of Stake (PoS) protocol \cite{kiayias2017ouroboros}. Therefore, to easily understand the applications of game theory in blockchain, the next section presents an overview and fundamentals of game models used in this survey.


\section{Overview and Fundamentals of Game Theory}\label{background}
Game theory provides a set of mathematical tools for analyzing the interaction among rational decision-makers. In a game, each decision-maker as a player chooses its strategy to maximize its utility, given the other players' strategies. The following briefly presents the game theoretic approaches which have been widely applied to analyze the interactions within the blockchain network. To interpret the definition of the game, some important terminologies are given below.

\begin{itemize}
\item Player: A player is a decision-maker in the game. In the blockchain, players can be miners, mining pools, or the blockchain users.
\item Utility: A utility, i.e., a payoff, an interest, or a revenue reflects the player's expected outcome.
\item Strategy: A player's strategy is a set of actions, choices or decisions that the player can perform to achieve its expected outcome. In general, the player's utility is determined based on not only the player's own strategy, but also the other players' strategies.
\item Rationality: A player is rational, i.e., self-interested, if the player always maximizes its own payoff.
\end{itemize}

\subsection{Non-cooperative Game}
In a non-cooperative game, the players do not cooperate by forming coalitions or by reaching agreements. In general, the term \textit{non-cooperative} does not imply that the players do not cooperate with each other, but it means that any cooperation which might arise must be with no communication of strategies among the players. In other words, the strategy that the player takes must be spontaneous, and each player is rational.

Consider a blockchain network in which miners as the players invest strategically in computational power to compete for a reward from mining successfully. The miners are rational and the non-cooperative game can be used to model the interaction among the miners. Assume that there are $N$ miners, i.e., players, and $P_i$ is a set of strategies of miner $i$, where $P=P_1 \times \dots \times P_N$ is the Cartesian product of the sets of individual strategies. Let $p_i \in P_i$ be the strategy of miner $i$. A vector of strategies of $N$ miner is $\mathbf{p}=(p_1,\dots,p_N)$, and a vector of corresponding payoffs is $ \boldsymbol{\pi}=(\pi_1(\mathbf{p}),\dots,\pi_N(\mathbf{p})) \in R^N$, where $\pi_i(\mathbf{p})$ is the utility of player $i$, e.g., mining rewards or the transaction fees, given the miner's chosen strategy and strategies of the others. Each miner chooses its best strategy $p_i^*$ to maximize its utility. A set of strategies $\mathbf{p^*}=(p_1^*,\dots,p_N^*) \in P$ is the Nash equilibrium if no miner can gain higher utility by changing its own strategy when the strategies of the other miners remain unchanged, i.e.,
\begin{equation}
\label{Nash_equilibrium}
\forall i, p_i \in P_i : \pi_i(p_i^*,\mathbf{ \overline{p}_i^*}) \geq \pi_i(p_i, \mathbf{\overline{p}_i^*}),
\end{equation}
where $\mathbf{\overline{p}_i}=(p_1,\dots,p_{i-1}, p_{i+1},\dots, p_N)$ is a vector of strategy of all miners except miner $i$.

The inequality in (\ref{Nash_equilibrium}) demonstrates the equilibrium state of the game. At the Nash equilibrium, the players have no incentive to deviate from their current strategies. However, there is no Nash equilibrium in some cases, or multiple equilibria exist. Thus, it is important to check the existence and uniqueness of the Nash equilibrium to analyze a non-cooperative game. The existence and uniqueness of equilibrium theory \cite{rosen1965existence} demonstrates that the strictly concave game can achieve the unique equilibrium asymptotically. Here, the concave game means that the utility functions of players are concave, and this can be proved by computing the second-order derivative of the utility function \cite{han2012game}.

The non-cooperative theory can be applied to a broad range of blockchain based scenarios. For example, it can be used for computational power allocation \cite{dimitri2017bitcoin} or fork chain selection \cite{carlsten2016instability}. Also, it can be used for pool selection regarding the mining rewards allocation \cite{schrijvers2016incentive}. Atop the blockchain based platform, the non-cooperative game theory is applied to analyze the interaction between blockchain users and miners, e.g., cheating among the buyers and sellers in blockchain network \cite{bigi2015validation}. Moreover, it is widely adopted in analysis of security issues within the blockchain, e.g., pool block withholding attacks  \cite{eyal2015miner}.

\subsection{Extensive-form Game}\label{spe}

\begin{figure}[htbp]
 \centering
\includegraphics[width=5.8 cm, height=4cm]{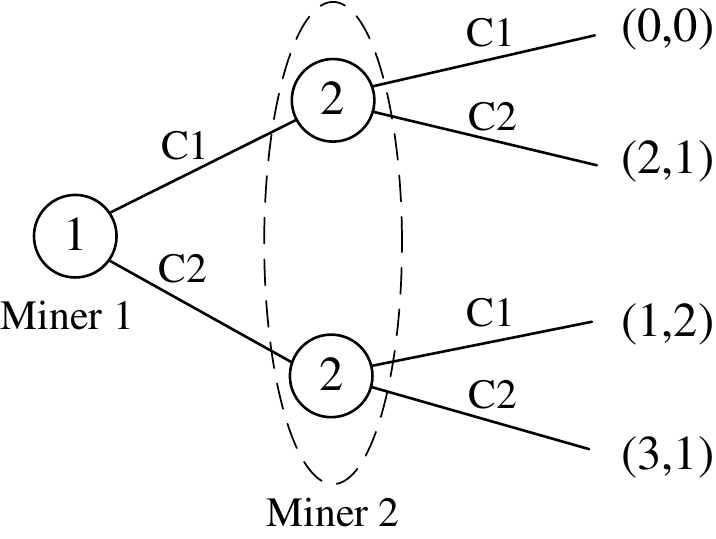}
 \caption{The game has two players, i.e., miner 1 and miner 2. The initial node belongs to miner 1 meaning that the miner 1 makes its strategy first. The miner 1 chooses between Chain 1, i.e., C1, and Chain 2, i.e., C2. The miner 2 chooses between C1 and C2 after its observation of the action of miner 1. There are four payoffs represented by the four terminal nodes of the game tree: (C1,C1), (C1,C2), (C2,C1) and (C2,C2).}
 \label{gametree}
\end{figure}

The aforementioned non-cooperative game can be used to analyze both the static game, i.e., the game that has no notion of time and no player has any knowledge of other players' actions in advance, and the dynamic game, i.e., the game in which the players' strategies are made following a certain predefined order. The dynamic game can be represented in an extensive form to illustrate the sequencing of players' possible moves, their choices at every decision point, information that each player has about the other players' moves, and their payoffs for all possible game outcomes. In game theory, the extensive-form game describes the interaction among the players using a game tree illustrating decisions made at different points with their payoffs represented at the end of each branch. Consider the scenario of fork chain selection, the miner chooses a certain chain to mine on at the beginning of every round of mining competition, given the actions taken by the other players in previous mining rounds. At some points, the blockchain forks and leads to the structure similar to a branching tree. Thus, the tree-like extensive-form game can be efficiently applied for the analysis as shown in Fig. \ref{gametree} in which the players can choose between two chains to mine.

%

Assume that an extensive-form game is composed of many smaller games, i.e., subgames. Each subgame can be expressed as a static non-cooperative game. A set of strategies $\mathbf{p^*}=(p_1^*,\dots,p_N^*) \in P$ is a subgame perfect equilibrium if it represents a Nash equilibrium of every subgame. A common method for obtaining the subgame perfect equilibrium in an extensive-form game is backward induction. The backward induction first considers the decision that might be made in the last move and then reasons back from the end of the problem to the previous one until the induction reaches the first move of the game. In the game as presented in Fig. \ref{gametree}, if miner 1 chooses C2, miner 2 will choose C1 to maximize its utility and miner 1 receives 1. If miner 1 chooses C1, miner 2 will choose C2 and miner 1 receives 2. Therefore, miner 1 prefers choosing C1 and miner 2 choosing C2. The strategies of miners are the  Nash equilibrium of each subgame and thus achieve the subgame perfect equilibrium.

In blockchain based platform, the extensive-form game is applied for selection of entering the blockchain market or not \cite{cong2018tokenomics}, selection of transactions to be included in the block \cite{lewenberg2015inclusive}, and optimization of pool's mining rewards allocation \cite{cong2018decentralized}. The extensive-form game has been also adopted for analyzing the security issues within the blockchain. It was used to analyze the selection of fork chain \cite{saleh2018blockchain}, determination of forming the collusion \cite{dong2017betrayal}, and cheating among the blockchain users \cite{asgaonkar2018solving}.

\subsection{Stackelberg Game}

Similar to the extensive-form game, another game that involves in a certain predefined ordered strategies taken by players is \textit{Stackelberg game} \cite{han2012game}. In the Stakelberg game, the players include \textit{leaders} and \textit{followers}. The followers decide their strategies after observing the strategies of the leaders. Both the leaders and the followers are typically rational that aim to maximize their own utilities.

To understand how the Stackelberg game works, we consider a blockchain based edge computing network which involves two players, i.e., the service provider and the miner \cite{xiong2018mobile}. The service provider possesses the computational power which can be offered to the miner as service, and the provider can set the service price to charge the fee for profit. The miner optimizes its demand of computational power to the provider to maximize its utility, taking its cost into account. As such, the service provider sets the price first, and then the miner decides its demand. Thus, the Stackelberg game can be used to model the interaction between the service provider and the miner. Assume $P_1$ and $P_2$ are the sets of strategies of the service provider and the miner, respectively. The service provider chooses its strategy $p_1$ from set $P_1$ to maximize its utility $\pi_1(p_1,p_2)$, and the miner chooses its strategy $p_2$ from set $P_2$ to maximize its utility $\pi_2(p_1,p_2)$. The optimization problems of the leader and the follower together form the Stackelberg game. The objective of such a game is to find a Stackelberg equilibrium.

\textbf{Definition 1.} Let $\mathop{BR}_2(p_{1})$ define the best response mapping of the follower. Then, the point $(p_1^*, p_2^*)$ is called the Stackelberg equilibrium of the game if the following conditions hold:

\begin{itemize}
\item $p_2^*\in \mathop{\textrm{BR}}_2(p_1^*)$, and
\item $p_1^* \in \arg\max\limits_{p_1}\max\limits_{p_2\in\mathop{\textrm{BR}}_2(p_1)} \pi(p_1, p_2)$.
\end{itemize}

To find the Stackelberg equilibrium, the backward induction method is typically used. Since the leader first takes its strategy and then the follower chooses its strategy, the Stackelberg strategy guarantees the service provider to achieve its payoff at least as much as the corresponding Nash equilibrium.  The reason is that when choosing the Stackelberg strategy, the service provider actually optimizes its decision which will maximize its utility. This feature makes the Stackelberg game suitable for many scenarios in blockchain based applications. For example, the Stackelberg game is adopted for setting transaction fees and selection of miners for verification \cite{kangwcl}, determination of cyber-insurance price \cite{feng2018cyber}, and analyzing the supply-demand relationship in the blockchain based edge computing platform \cite{xiong2018optimal}.

\subsection{Stochastic Game}

A stochastic game can be seen as several static non-cooperative games that are repeated over time. Each static non-cooperative game is called \textit{state} of the game. The stochastic game executes \textit{stochastic transitions} among the states of the game. In the stochastic game, the players can change their strategies based on the past actions and transitions behaviors of the other players \cite{shapley1953stochastic}.

The stochastic game can be applied efficiently to analyze the miners' selection of chains to mine (see Section \ref{blkoverview}) regarding the transitions of blockchain structure. The stochastic game typically is composed of (i) a finite set $I$ of players , e.g., the miners, (ii) a space $M$ of states, e.g., blockchain structures, (iii) a strategy set $S$, and (iv) a transition probability $P$ from $M \times S$. Each miner has a payoff function $g_n$ which is often taken to be the discounted sum of the stage payoffs. The game starts at an initial state $m_1$, and at stage $t$, each miner observes the blockchain structure $m_t$ and then chooses its strategy $s_t^i$, i.e., selects a chain to mine. Every miner receives an immediate payoff $g_n^i$ associated with the current state and the miners' strategies. Then, the game moves to a new state $m_{t+1}$. The game process is repeated until it reaches a common solution called Markov Perfect Equilibrium (MPE) \cite{maskin2001markov} that is the refinement of the subgame perfect equilibrium (see Section III-B). The Markov perfect equilibrium is a set of strategies that achieve the Nash equilibrium of every state of the stochastic game \cite{han2012game}. In the case of fork chain selection, following the Nakamoto protocol, i.e., mining on the longest chain, is the Markov equilibrium.

Apart from the chain selection, the stochastic game can be used for mining management. For example, the selection between investing in computational power or leaving the mining \cite{dhamal2018stochastic}, and the selection of chain to mine \cite{biais2018blockchain}. Moreover, stochastic game has also been widely applied to security issues. It was used to analyze the selection between honest mining and selfish mining \cite{zhen2017zero}, the decision of the proper time to release the mined block \cite{kroll2013economics}, and the selection of adding a block to the chain \cite{kim2018trailer}.

\section{Applications of Game Theory for Security}

\subsection{Selfish Mining Attack}\label{SMA}

Selfish mining is a type of subversive strategies in PoW based blockchain systems \cite{eyal2018majority} that attackers, i.e., malicious miners or mining pools, may not broadcast the newly mined blocks but choose to (i) withhold the block or (ii) hold and then release the block at a proper time. Under this case, honest miners waste their computational power in finding the block discovered already, and malicious miners can therby increase their probability of finding the next block. The pool block withholding (PBWH) attack is one common selfish mining attack \cite{courtois2014subversive}. In the PBWH attack, the attacking pool infiltrates the attacked pool, and the infiltrating miners perform the block withholding (BWH) attack, i.e., withhold all the blocks newly discovered in the attacked pool. To prevent such an attack, it is crucial to analyze strategies of the miners and pools as well as the interaction among them. A Markov Decision Process (MDP) \cite{sapirshtein2016optimal} can be used to analyze the strategy and utility of the individual player, i.e., the miner or the pool. However, the MDP does not take into account the interaction among multiple players. Alternatively, game theory can be effectively applied.

\begin{table*}
\caption{\label{tab: }A Summary of Game Theoretical Applications for Security.}
\label{Security_summary_table}
\begin{centering}
\begin{tabular}{|>{\centering\arraybackslash}m{0.3cm}|>{\centering}m{0.5cm}|>{\centering\arraybackslash}m{1.9cm}|>{\centering}m{2cm}|>{\centering}m{2.4cm}|>{\centering}m{3.5cm}|>{\centering}m{2.2cm}|>{\centering}m{2cm}|}
\hline
 \cellcolor{mygray}  \textbf{\noun{}} &  \cellcolor{mygray} \textbf{\noun{Ref.}}&  \cellcolor{mygray} \textbf{\noun{Game Model}}&  \cellcolor{mygray} \textbf{\noun{Player}}&  \cellcolor{mygray} \textbf{\noun{Action}}  &  \cellcolor{mygray} \textbf{\noun{Strategy}} &  \cellcolor{mygray}\textbf{\noun{Payoff}} & \cellcolor{mygray} \textbf{\noun{Solution}} \tabularnewline
\hline
\hline

\parbox[t]{2mm}{\multirow{6}{*}{\rotatebox[origin=c]{90}{ \hspace{-5 cm} Selfish Mining Attack}}}
&\cite{eyal2015miner}& Non-cooperative game &Mining pools& Infiltrate other pools to launch BWH attack &Determination of the infiltration rate& Mining rewards minus cost & Nash equilibrium \tabularnewline \cline{2-8}
&\cite{luu2015power}& Splitting game &One miner and pools & Distribute mining power for selfish mining &Determination of the power distribution&Mining rewards minus cost & Mixed strategy Nash equilibrium  \tabularnewline \cline{2-8}
&\cite{chatterjee2018ergodic}& Mean-payoff game & Mining pools & Migrate to other pools to launch PBWH attack & Determination of the migration rate& Mean-payoff & Mean-payoff objective \tabularnewline \cline{2-8}
& \cite{zhen2017zero}& Stochastic game & Miners & Block withholding (BWH) attack & Selection between honest mining and selfish mining & Social welfare & Zero-Determinant strategy \tabularnewline \cline{2-8}
&\cite{babaioff2012bitcoin}& Non-cooperative game & Miners & Selfish propagation attack & Selection of identity duplication and transactions relaying & Mining rewards & Nash equilibrium \tabularnewline \cline{2-8}
&\cite{carlsten2016instability}& Non-cooperative game & Miners & Fork chain & Selection of fork to mine & Transaction fees & Nash equilibrium \tabularnewline \cline{2-8}
&\cite{zolotavkin2017incentive}& Non-cooperative game & Miners & Delay submitting shares & Decision of the proper time to submit shares & Mining rewards & Nash equilibrium \tabularnewline \cline{2-8}
&\cite{kroll2013economics}& Non-cooperative game & Miners & Select or create a chain to mine & Selection of the chain to mine & Mining rewards & Nash equilibrium \tabularnewline \cline{2-8}
\hline

\parbox[t]{2mm}{\multirow{12}{*}{\rotatebox[origin=c]{90}{ \hspace{-3 cm} majority Attack}}}
&\cite{kroll2013economics}& Stochastic game & Miners & BWH attack & Decision of the proper time to release the block & Mining rewards & Nash equilibrium \tabularnewline \cline{2-8}
&\cite{teutsch2016cryptocurrencies}& Non-cooperative game & Miners & Post smart contract transaction of mining on private chain & Selection between working on smart contract transaction and honestly mining &Transaction fees and mining rewards&Nash equilibrium \tabularnewline \cline{2-8}
&\cite{kim2018trailer}& Stochastic game & Miners & Compete to fork chain & Selection of adding the block to the chain &  Mining rewards minus cost & Nash equilibrium\tabularnewline \cline{2-8}
&\cite{liao2017incentivizing}& Non-cooperative game & Attacking and defending miners & Issue whale transaction to attract miners mine on the private chain & Determination of the threshold of attack cost and block selection & Mining reward minus cost & Nash equilibrium \tabularnewline \cline{2-8}
&\cite{houy2014will} & Sequential game & Attacking and defending miners & Buy stake to launch majority attack & Determine the cost of attack and selling selection & Function of profit and interest & Nash equilibrium\tabularnewline \cline{2-8}
&\cite{kroll2013economics}& Non-cooperative game & Attacking and defending miners & Goldfinger attack & Decision of forming cartel and determination of the tax paid to the attacker & Profits minus cost & Nash equilibrium \tabularnewline \cline{2-8}
&\cite{kangwcl}& Stackelberg game & Blockchain users and miners & Form cartel to launch majority attack & Setting transaction fee and selection of recruiting miners & Profits minus cost & Stackelberg equilibrium \tabularnewline \cline{2-8}
\hline

\parbox[t]{2mm}{\multirow{1}{*}{\rotatebox[origin=c]{90}{ \hspace{-3 cm} DoS Attack}}}
&\cite{johnson2014game}& Non-cooperative game & Mining pools & DDoS attack & Selection of launching attack or not & Profits minus cost&Nash equilibrium \tabularnewline \cline{2-8}
&\cite{laszka2015bitcoin}& Sequential game & Mining pools & DDoS attack & Chosen of the attack level & Profits minus cost & Nash equilibrium\tabularnewline \cline{2-8}
&\cite{nojoumian2018incentivizing}& Repeated game & Mining pools & DDoS attack under a reputation-based scheme & Selection of launching attack or not &Profits associate with the loss of reputation&Nash equilibrium \tabularnewline \cline{2-8}
&\cite{xugame}& Non-cooperative game & One server and devices & DDoS attack in edge network & Selection between executing or sending request and launching attack &Profits minus cost& Nash equilibrium \tabularnewline \cline{2-8}
\hline

\parbox[t]{2mm}{\multirow{10}{*}{\rotatebox[origin=c]{90}{ \hspace{-3 cm} Other security issues}}}
&\cite{rawat2018ishare}& Non-cooperative game & Groups of information sharing network & Form group and infiltrate other groups to withhold data & Determination of infiltration rate & Profits minus cost & Nash equilibrium \tabularnewline \cline{2-8}
&\cite{dong2017betrayal}& Extensive-form game & Clouds of cloud computing network & Collude to output the same wrong data & Selection of collusion or not & Function of payment and deposit&Sequential equilibrium \tabularnewline \cline{2-8}
&\cite{asgaonkar2018solving}& Extensive-form game & Buyer and seller of the blockchain trading system & Cheats of buyer or seller & Selection of cheating or not &Profits associated with deposits&Subgame perfect Nash equilibrium \tabularnewline \cline{2-8}
&\cite{bigi2015validation}& Non-cooperative game & Buyer and seller of the blockchain trading system & Cheats of buyer or seller & Selection of cheating or not &Profits associated with deposits & Nash equilibrium \tabularnewline \cline{2-8}
&\cite{adler2018astraea}& Coordination game & Voter and verifiers& Manipulate data  of data verification system & Statement of the correctness of data & Profits associated with deposits & Nash equilibrium \tabularnewline \cline{2-8}
&\cite{feng2018cyber}& Stackelberg game & Blockchain users, one provider, and one insurer & Purchase insurance to compensate for the attack & Determination of the service price, service demand, and insurance price & Profits minus cost & Stackelberg equilibrium \tabularnewline \cline{2-8}
\hline

\end{tabular}
\par\end{centering}
\end{table*}

\begin{figure}[htbp]
 \centering
\includegraphics[width=5.8 cm, height=2.5cm]{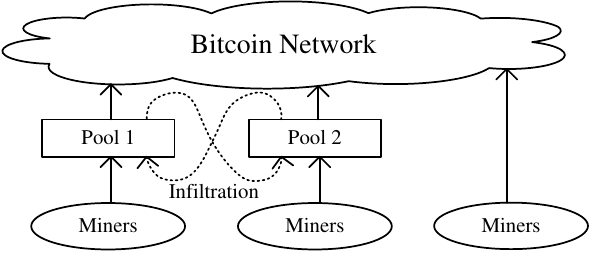}
 \caption{Two pools case that both pools launch the PBWH attack, i.e., infiltrates each other with its miners that perform the BWH attack \cite{eyal2015miner}. }
 \label{twopoolattack}
\end{figure}

The authors in \cite{eyal2015miner} adopt a non-cooperative game to analyze the interaction among the pools. This scenario is illustrated in Fig. \ref{twopoolattack} with two selfish pools as players. The strategy of each player is to determine its infiltration rate, i.e., the fraction of its computational power for performing the infiltration. In the case of attack, the attacking pool obtains its utility not only from its honest miners, but also from the infiltrating miners that perform the BWH attack within the attacked pool. The objective of the player is to optimize its infiltration rate thereby maximizing its utility. In particular, the player's utility is a function of the computational power and the infiltration rate. By using the second-order derivative with respect to the infiltration rate, the utility function is proved to be concave. Thus, there exists a unique Nash equilibrium in which neither players can improve its own utility by changing its strategy, i.e., the infiltrate rate. At the equilibrium, the infiltrate rate is always greater than zero. This means that launching the PBWH attack is always the best response of each player. Simulation results illustrate that the pool can improve its utility by launching the PBWH attack only when the pool controls a strict majority of the total computational power. However, in the case that two pools attack with each other, the utility of each pool is less than that if neither pool attacks.

The case in \cite{eyal2015miner} is similar to the famous Prisoners' Dilemma in game theory \cite{han2012game} that results in the utility loss of the miners. To avoid the miners' dilemma, the miners can choose one of the solutions as follows. The first solution is that the miners would intend to join private pools that will not involve the PWH attack. As a result, big mining pools may be divided into many small pools spontaneously, and eventually this may lead to a better environment for the Bitcoin system as a whole. The second solution is that the miners perform so-called Zero Determinant (ZD) strategies \cite{he2016zero}. This solution is presented in \cite{zhen2017zero} that the authors model a two-miner mining case as a stochastic iterative game.

Different from a typical strategy that aims to improve players' own profits, the ZD strategy is used to control an outcome of the opponents in a certain range so as to avoid a low social welfare, i.e., the whole pool's profit \cite{zhang2016zero}. In this game, the two players are an altruistic miner, i.e., a miner which attempts to maximize the social welfare, and a selfish miner, i.e., a miner which only aims to improve its own profit. Their strategies include cooperation, i.e., mining honestly, and launching the BWH attack to the other miner. Note that the altruistic miner and selfish miner choose their strategies probabilistically based on each other's strategy selected in the last iteration. The analysis shows that so long as the altruistic miner applies strategies according to the determinant function, i.e., a linear function which is associated with players' profit factor, the profit of the selfish miner is in a range from mutual cooperation to mutual attack regardless of strategies adopted by the selfish miner. Thus, the altruistic miner can indeed motivate the selfish miner to mine cooperatively by performing ZD strategies so as to restrict the selfish miner's profit to achieve the highest social welfare. The simulation results show that the proposed game can achieve a higher social welfare than that of the pool game proposed in \cite{eyal2015miner}. However, the proposed game does not consider the profit of the altruistic miner. This means that the altruistic miner may not have an incentive to perform the ZD strategy.

The two-pool-attacker scenario in \cite{zhen2017zero} can also be found in \cite{chatterjee2018ergodic}. However, in addition to the PBWH attack, the authors in \cite{chatterjee2018ergodic} consider the miners' migration among the pools. In particular, the miners of a pool can be migrated to another pool and launch the PBWH attack to increase the profit. To analyze the average payoff of the miner and the miners' stochastic migration process, the Concurrent Mean-payoff Game (CMPG) is adopted as presented in \cite{chatterjee2018ergodic}. CMPG (see Section \ref{background}) is a two-player game with a finite state space where at each state, both players choose their strategies simultaneously \cite{shapley1953stochastic}. Here, the players are pool $1$ and pool $2$, and the state of the game includes the number of migrated miners of pool $1$ and that of pool $2$. The strategy of a pool is to determine (i) the number of its miners to be migrated to the other pool and (ii) the miners which perform the PBWH attack. The number of migrated miners is determined depending on the attractiveness levels of the other pool, i.e., the ratio of the pool's total mining reward to the total computational power of its miners. If a pool is infiltrated by miners of the other pool, the attractiveness level of the pool decreases. This decrease can be observed by the whole blockchain network, and thus the other pool can adjust its migration strategy based on the observations. In general, the pool's profit depends not only on the state, i.e., the allocation of miners for migration, but also on its chosen strategy. The experimental results show that if the miners in pool 1 stochastically migrate to pool 2 according to the pool 2's attractiveness level, then the mean-payoff objective, i.e., the average profit, of pool $2$ can be guaranteed against any strategy of pool $1$. However, the mean-payoff objective may not be guaranteed in multi-player scenarios. Such a scenario can be investigated in the future work.

\begin{figure}[htbp]
 \centering
\includegraphics[width=5.5 cm, height=3.2cm]{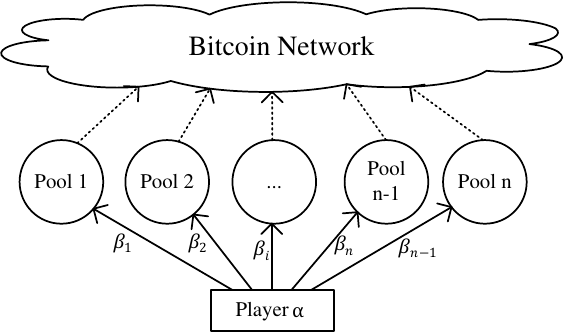}
 \caption{ The player $\alpha$ distributes its computational power to several pools and launchs the BWH attack, where $\alpha$ is the computational power owned by the player, and $\beta_i$ represents the fraction of mining power that the player allocates to the pool $i$ \cite{luu2015power}. }
 \label{CPS}
\end{figure}

The aforementioned approaches, i.e., \cite{eyal2015miner}, \cite{zhen2017zero} and \cite{chatterjee2018ergodic}, are constrained to the interaction among only two pools. Considering a multi-pool scenario, the authors in \cite{luu2015power} adopt the Computational Power Splitting (CPS) game \cite{laraki2002splitting} to model the PBWH attack. To improve their expected payoffs, the players, i.e., the miners or the pools which own positive computational power, can choose to (i) attack other pools, i.e., distribute their computational power to other pools and launch the BWH attack, and (ii) honestly follow or arbitrarily deviate from the pool's protocol. In the case that the player chooses to attack, the strategy of the player is to determine (i) the distribution of its computational power, and (ii) the portion of its mining power holding attack as presented in Fig. \ref{CPS}. The objective is to maximize the player's profit, which is defined as the sum of mining rewards received from all the pools. For any given strategies of the other miners, there always exists a computational power allocation for a miner to increase its profit and cause the other pool a loss. In other words, honestly mining is not the best response of the players and the game thus has no pure Nash Equilibrium strategy. Nonetheless, the game has a unique mixed strategy equilibrium at which each player has an incentive to launch the PBWH attack probabilistically rather than mining honestly. Simulation result shows that the best strategy of the players is to comply with the following rules. First, the players launch the PBWH attack which improves their profits. Second, the attackers spend the computational power less than a specific fraction on the PBWH attack to gain more profit than mining honestly. Finally, the attackers should attack big pools rather than small pools. Both work in \cite{eyal2015miner} and \cite{luu2015power} arrive at some consistent findings from different perspective.

\begin{figure}[htbp]
 \centering
\includegraphics[width=8 cm, height=3.4cm]{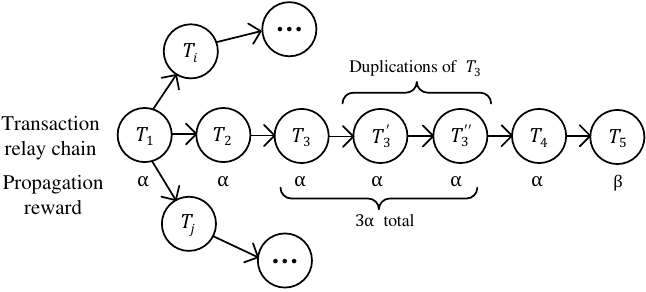}
 \caption{ An example of the transaction relay process that the transaction flows from $T_1$ to $T_5$. $T_1$ to $T_4$ relay the transaction thus gain reward $\alpha$. $T_5$ solves the PoW thus gains reward $\beta$. $T_3$ adds two fake identities, i.e., $T_3^{'}$ and $T_3^{''}$, before relaying the transaction thereby gains $3\alpha$ in total \cite{babaioff2012bitcoin}. }
 \label{duplication}
\end{figure}

The approaches discussed above, i.e., \cite{eyal2015miner} and \cite{luu2015power}, consider only the mining reward. In practice, the Bitcoin systems also provide the transaction fee \cite{nakamoto2008bitcoin}. When the block creation reward dominates the mining reward, the miners may not broadcast transactions to the others immediately so as to increase their expected profits \cite{decker2013information}. This is called \textit{selfish propagation attack}. To address the attack, the authors in \cite{babaioff2012bitcoin} propose an incentive mechanism for the miners to propagate the transactions. The proposed mechanism is designed such that each miner receives a propagation reward from the blockchain system according to its behaviors in the propagation process (see Section \ref{property}). To maximize the gained propagation reward, each miner strategically chooses to duplicate itself, i.e., add fake identities before relaying the transaction, or to relay the transaction immediately, given the strategy profile of the other miners. The interaction among the miners can be modeled as a non-cooperative game as presented in \cite{babaioff2012bitcoin}. In the game, the players are miners which are aware of the transaction. Each player not only strategically relays the transaction but also works on PoW. The authorizing player, i.e., the player which solves the PoW, and the players which are in the same relay chain with the authorizing player gain a certain reward. Other players gain nothing. This scenario is illustrated in Fig. \ref{duplication}. By using the iterative removal of dominated strategies \cite{han2012game}, the game is proved to admit a unique Nash equilibrium. At the Nash equilibrium, only transaction propagating and no-duplication strategies, i.e., the Nash equilibrium strategy, survive after dominated strategy removal. However, if there are not sufficient players which are connected with each other, the selfish propagation attack cannot be guaranteed to be prevented.

\begin{figure}[htbp]
 \centering
\includegraphics[width=8 cm, height=3.5cm]{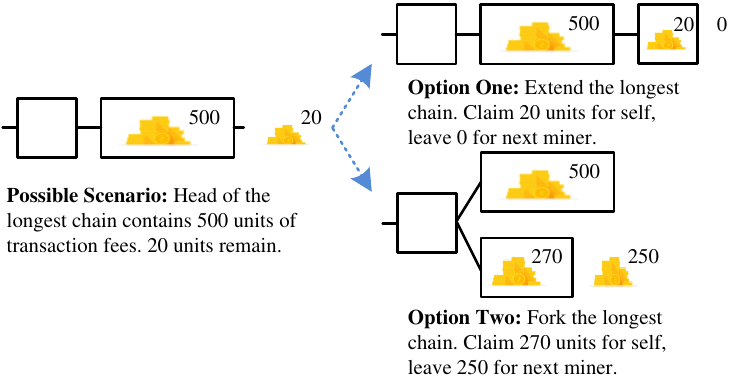}
 \caption{An example of undercutting attack. Option one corresponds to honest mining that the miner mines on the longest chain. The miner in Option Two performs undercutting attack that forks the longest chain and claims more reward compared with that of option one \cite{carlsten2016instability}. }
 \label{undercutting}
\end{figure}

Other works on understanding the vulnerability of propagation mechanism without mining rewards can also be found in \cite{kroll2013economics,houy2014economics,rizun2015transaction}. The authors in \cite{carlsten2016instability} demonstrate that with only block creation rewards, it is attractive enough for miners to extend the blocks that have the most available transaction fees rather than to follow the longest chain. Each miner intends to fork the head of the chain actively and leaves transactions unclaimed selectively to maximize its profit. Such an attack is called \textit{undercutting attack}, and the miner that performs the undercutting attack is called \textit{undercutter}. The scenario is illustrated as in Fig. \ref{undercutting} where ``Option Two'' corresponds to the undercutting attack. If the miner performs the undercutting strategy, it may gain nothing if its block is not in the longest chain eventually. The undercutter strategically performs undercutting strategy so as to attract the other miners to mine on the forked chain. Meanwhile, the other miners consider whether to mine on the forked chain or not to maximize their profits. Thus, the interaction among the miners can be modeled as a repeated game that in every stage of mining, each miner chooses to perform honest mining or undercutting. The game theoretical analysis shows that if a miner's undercutting strategy follows a certain function to maximize the size of the block, then the strategy is also the best response for all miners. This is under the constraint that if the miners fork, they must perform undercutting. Thus, the Nash equilibrium exists as all miners adopt the same undercutting strategy. The simulation results show that when each miner applies a no-regret learning algorithm, even with 66\% of miners mining honestly, undercutting is profitable than mining honestly. As a result, there could be many unclaimed transactions left which is detrimental to the whole blockchain network. The same conclusion is reached in \cite{moser2015trends} through a non-game theoretical method. However, if the simulation takes network latency into account, the undercutters may have sufficient time to include all the transactions into the block, and thus the undercutters have no incentive to leave any transaction to the next miner.

\begin{figure}[htbp]
 \centering
\includegraphics[width=8 cm, height=3.3cm]{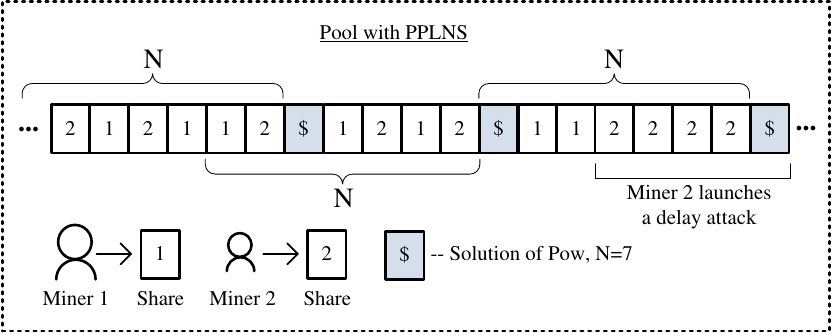}
 \caption{ An example of delay attack in a pool with PPLNS. The pool includes two miners, i.e., miner 1 and miner 2. The size of PPLNS window in this case is 7. Miner $2$ launches a delay attack \cite{zolotavkin2017incentive}.}
 \label{PPLNS}
\end{figure}

Different from attacks among the pools, another variation of selfish mining attack inside the mining pool which performs on the pay per last N shares (PPLNS) \cite{rosenfeld2011analysis} is introduced in \cite{zolotavkin2017incentive}. PPLNS is a popular pool mining reward mechanism. Instead of distributing a block reward among miners in the pool in the current round, PPLNS distributes the reward among miners that have submitted shares\footnote{ A share is a hash value which is easier to be found, compared with the valid hash of the block. This means that shares can be used to measure the computational power that miner possesses.} already in the latest PPLNS window. The PPLNS window includes the number of shares submitted continuously that the latest share is the full solution of PoW. Specifically, shares in the PPLNS window are regarded as the effective shares. The miner that submits effective shares obtains the reward according to its proportion of all effective shares. Under this mechanism, the miner may launch the delay attack. In the delay attack, the miner first delays submitting the shares, i.e., by holding the discovered shares, if the miner finds the solution of PoW, the miner releases all delayed shares and then submits the solution immediately. Thus, more reward can be obtained because of the higher fraction of shares in the latest PPLNS window. This scenario is illustrated in Fig \ref{PPLNS}.

For each miner in the same pool, there are two phases during mining. In the first phase, the miner only collects shares for delay. In the second phase, the miner submits every share immediately, i.e., through honest mining. To maximize the expected profit of launching the delay attack, each miner needs to choose proper time to transit its phase according to the strategies of the other miner. Otherwise, the miner may lose the reward of all its delayed shares. Thus, the authors in \cite{zolotavkin2017incentive} model the interaction between miners in the same pool as a non-cooperative game. It is proved that the Nash equilibrium exists if the computational power of the most powerful miner meets a certain condition. This condition is associated with the PPLNS window size, and complexity of finding the solution of PoW. At the Nash equilibrium, each miner of the pool is at the turning point between two phases. This means that the miner has no incentive to deviate from honest mining, and thus the miner would not delay its shares. Such a pool is called the \textit{incentive compatible} pool. Simulation results show that if the pool is not incentive compatible, although the fraction of delaying miners decreases with a parameter related to the window size and the complexity of solving PoW, regardless of power distribution, the game cannot reach the Nash equilibrium.

\subsection{Majority Attack}

The security of blockchain is achieved through the distributed consensus of miners. This consensus is only reliable with the assumption that no single miner can hold more than 50\% of the network's computational power \cite{nakamoto2008bitcoin}. Theoretically, to gain its profit, the miner invests more in the computational power, and it may possess more than 50\% of the network's computational power \cite{schrijvers2016incentive}. In this case, the miner would be able to halt payments, reverse transactions, prevent new transactions from confirmation, and double-spend coins \cite{bradbury2013problem,li2017survey,zheng2016blockchain,barber2012bitter,akcora2017blockchain,bentov2016cryptocurrencies,rosenfeld2014analysis,shomer2014phase}. The attack is called 51\% attack. As such, the assumption of the distributed consensus may not be valid any longer, and the security of blockchain is not guaranteed. More specifically, theoretical analyses \cite{bahack2013theoretical,sapirshtein2016optimal,teutsch2016cryptocurrencies} show that the miner which possesses only a relatively large part computational power can also achieve the similar goal. In general, we label this type of attack associated with a large group of miners as the majority attack.

When the majority attack is performed, mining on the fork chain may happen. The condition under which a miner has an incentive to mine on the fork is investigated in \cite{kroll2013economics}. Although the miners follow the longest chain rule under the Nakamoto protocol, at some points, the chain can fork that leads to a structure similar to a branching tree \cite{courtois2014longest}. To maximize the profit, i.e., the reward of creation of a new block, each miner aims to extend selectively any of the existing branches or to create a new branch, given the strategy of the other miners. A non-cooperative game can thus be applied. Since if more than 50\% of the network's total computational power are extending the longest chain, deviating from honest mining only leads to the waste of the miner's computational power of mining. The reason is that the mined block would not achieve the Nakamoto consensus with the majority of miners and thereby be orphaned. This lowers the miner's profit, and thus following the longest chain would be the best response of the rest of other miners. Therefore, the game has a Nash equilibrium in which all miners extend the longest chain. If a cartel of miners which possesses more than 50\% of the network's computational power forks a chain, following the rule of longest chain would not be the best response for the other non-cartel miners, and thus the Nash equilibrium will be shifted to another one that every miner mines on the fork. Similar conclusion is reached in \cite{bahack2013theoretical}. If the fraction of computational power deviating from extending the longest chain is more than a value around $1/4$, each miner has an incentive to mine on the fork.

Compared with \cite{kroll2013economics}, a more general majority attack where the miner can not only choose which branches to mine but also determine whether or not to release the mined block is investigated in \cite{kiayias2016blockchain}. The miner can probabilistically hide newly mined blocks and mine on the fork. Since this is similar to that the miners play a game with incomplete information of blockchain state among each other \cite{anceaume2017bitcoin}, a stochastic game can be applied as presented in \cite{kiayias2016blockchain}. The miner's expected utility is a function of the miner's action, i.e., the allocation of the miner's computational power, and the current state of the game, i.e., the structure of the block tree at present. In the case that the miner's computational power is equal to a profit threshold, the expected utility of mining on a fork is equal to that of mining on the longest chain regardless of the current state. Thereby, when the miner's computational power is less than the profit threshold, the miner has no incentive to deviate from mining on the longest chain which is the best response of the miner and the Nash equilibrium can be obtained. As shown in the simulation results, when the obtained profit threshold is approximately 0.42, the miner with at most 36\% of the total computational power cannot gain more than 36\% of the total rewards. Meanwhile, the miner with computational power more than 46\% always has an incentive to deviate from the longest chain rule. These results are more accurate than that obtained by MDP-based scheme \cite{sapirshtein2016optimal}.

Furthermore, by using the smart contract \cite{buterin2014next}, the authors in \cite{teutsch2016cryptocurrencies} illustrate that the miner or the pool which controls only 38.2\% of the network's total computational power can gain more reward by deviating from the protocol. The attacking miner uses its full computational power to mine on its private chain while posting a smart contract transaction. This contract transaction includes a hashing puzzle, i.e., the solution of PoW, of its private chain. Any miner that solves the puzzle can receive the reward from the puzzle's giver, i.e., the attacker, in exchange for the solution. Thereby, the attacker may gain more profit when its private chain is longer than the public one. For each time that the attacker posts a hashing puzzle through the smart contract, the other miners have two strategies: (i) work on the puzzle in the contract, and (ii) mine on the public chain. Each miner tries to maximize its expected utility, given the set of strategies of the other miners. The interaction among the miners except the attacker can thus be modeled as a non-cooperative game. When the attacker controls more than 38.2\% of the network's total computational power, the miner's utility of working on the puzzle with probability $\alpha$ is greater than that of the mining the longest chain, and the attack is thus launched successfully. This means that each miner will work on the puzzle with probability $\alpha$ and mine on the public chain with probability $1-\alpha$. Thus, the game is proved to admit a mixed strategy Nash equilibrium. The game in \cite{teutsch2016cryptocurrencies} is under the assumption that miners always mine on the longest chain. However, if some miners perform the selfish mining strategy, the reward of solving the hashing puzzle on a private chain provided by the attacker may not be attractive enough to the other miners. Thus, the attack may fail.

In addition to posting the smart contract as presented in \cite{teutsch2016cryptocurrencies}, majority attack can also be launched by the attackers offering monetary bribes \cite{bonneau2016buy}. To prolong the fork chain thereby increasing its successful attack probability, the attacker can attract other rational miners to mine on the fork by issuing a \textit{whale transaction}, i.e., a transaction with a high transaction fee. Since issuing the whale transaction is similar to bribing the other miners, such an attack is also called \textit{bribery attack} \cite{becker2013can}. The attacker's problem is to determine the cost of the attack, i.e., the transaction fee, to maximize its profit. Also, the other miners' problem is to trade off the profit of mining on the fork against the reward of mining on the public chain. A non-cooperative game can be thus used to model the interaction between the attacker and the other miners as presented in \cite{liao2017incentivizing}. Both theoretical analysis and simulation results show that if the attacker's mining power is greater than a profit threshold, the cost of the attack decreases, i.e., the attacker's profit increases, as the attacker's mining power increases. Here, the profit threshold is a function of the computational power used to mine on the fork, and the number of blocks by which the fork chain is ahead of the public chain. Meanwhile, any miner that possesses as much mining power as the attacker's has an incentive to mine on the fork chain. However, the Nash equilibrium of the game is not discussed.

To avoid such majority attack, the existing miners can act as a defender actively adding honest nodes to the blockchain network. This case is investigated in \cite{kim2018trailer}. The system model consists of one attacker, i.e., the miner which intends to fork a private chain, and one defender, i.e., the miner which honestly mines on the public chain. To obtain the mining rewards, the attacker and the defender compete to build the blocks for the private and public chains in a sequence of stages, respectively. The historical strategies and the probabilistic stage transitions can be observed by both the attacker and the defender. Thus, the interaction between the attacker and the defender can be modeled as a stochastic game. In the game, the strategies of the defender are (i) \textit{defending}, i.e., actively adding the honest nodes to avoid the majority attack, and (ii) \textit{doing nothing}, i.e., letting the blockchain network run as usual. If the winning probability of the attacker to fork successfully is greater than a certain value, the defender's utility of defending is greater than that of doing nothing. This means that the defending strategy is the best response of the defender and the game reaches the Nash equilibrium. Here, the value is a determined based on the cost of adding honest nodes, the number of nodes added actively to the blockchain network, and the total mining power that the attacker has. Otherwise, the defender has no incentive to avoid the attack. However, the simulation results should use actual data gathered from real blockchain networks, e.g., Blockr.io \cite{blockrio} and Blockchain.info \cite{blockchaininfo}, to verify the practicability of the game model.

The aforementioned approaches, i.e., \cite{kiayias2016blockchain,teutsch2016cryptocurrencies,liao2017incentivizing,kim2018trailer}, consider the motivation of the attack within the system. However, the attacker's motivation can be also based on the incentive outside the system and this type of attack is called \textit{Goldfinger attack} \cite{mccorry2018smart}. The attacker, i.e., miners, can receive some utility from devaluing the cryptocurrency (a.k.a. currency measured in digital tokens in the blockchain network), by forming a cartel to impair the consensus among miners, i.e., launch the majority attack. The defenders, i.e., the other miners, intend to preserve the value of the currency. To prevent the currency from being devalued, the defender makes a bid, i.e., similar to a tax to keep the currency alive, to the attacker. Meanwhile, the defender trades off the cost of making the bid and the profit of preserving the currency. Therefore, a non-cooperative game can be used to model the interaction between the attacker and the defender as presented in \cite{kroll2013economics}. The utility of the miner is a function of the value of the currency, the bid, and the possibility of the currency being attacked. The analysis shows that the defender can maximize its utility by using the first-order optimality condition in which the bid satisfies a certain constraint associated with the computational power distribution. If such a bid exists, the game is at the Nash equilibrium point where the attacker has no incentive to attack. Otherwise, the currency will have a zero value. However, in a real case, the defender does not know the attacker's expected utility. If the attacker makes a strong claim about the imminent attack, the defender has no incentive to preserve the currency because of the possible high cost and thus no equilibrium exists.

Apart from the PoW system, the majority attack happens in Proof-of-Stake (PoS) systems \cite{kiayias2017ouroboros}. In a PoS system, each agent, i.e., a stake-holder, can earn interest by holding crypto-currency (CC) units (see Section \ref{background}). To improve the interest, the agent can make a price offer to buy CC units from other agents \cite{king2017ppcoin}. As an agent possesses more than 50\% of CC units of the system, this agent can halt and reverse any payments or transactions. Thus, the consensus of the system is broken and CC loses its value. Only the agent that intends to devaluate the CC obtains the profit, e.g., law enforcement, outside the system. The attack is typically launched in multiple stages \cite{houy2014will}, and thus each agent, i.e., the attacker or one of other agents, can observe the historical strategies of each other and then optimize its own strategy. Therefore, a sequential game can be used to model the interaction between the attacker and other agents as proposed in \cite{houy2014will}. In the game, the players include one attacker and other agents. The attacker trades off the profit of devaluating the CC against the cost of making offers and the loss of interest. In the case that the profit of devaluating the CC is greater than the interest of holding the CC, by using the backward induction method, the game is proved to admit a unique Nash equilibrium. At the equilibrium, the attacker has an incentive to buy more than 50\% of CC units, and other agents are willing to sell the CC to the attacker since they know that the CC has no value. However, the attacker can succeed in its attack at no cost by announcing to other agents about launching the majority attack before making the price offer. The reason is that if the agents believe that the attack succeeds, they will sell the CC to the attacker regardless of the price that attacker offers. The Nash equilibrium may not exist in this case.

\begin{figure}[htbp]
 \centering
\includegraphics[width=5.8 cm, height=3.3cm]{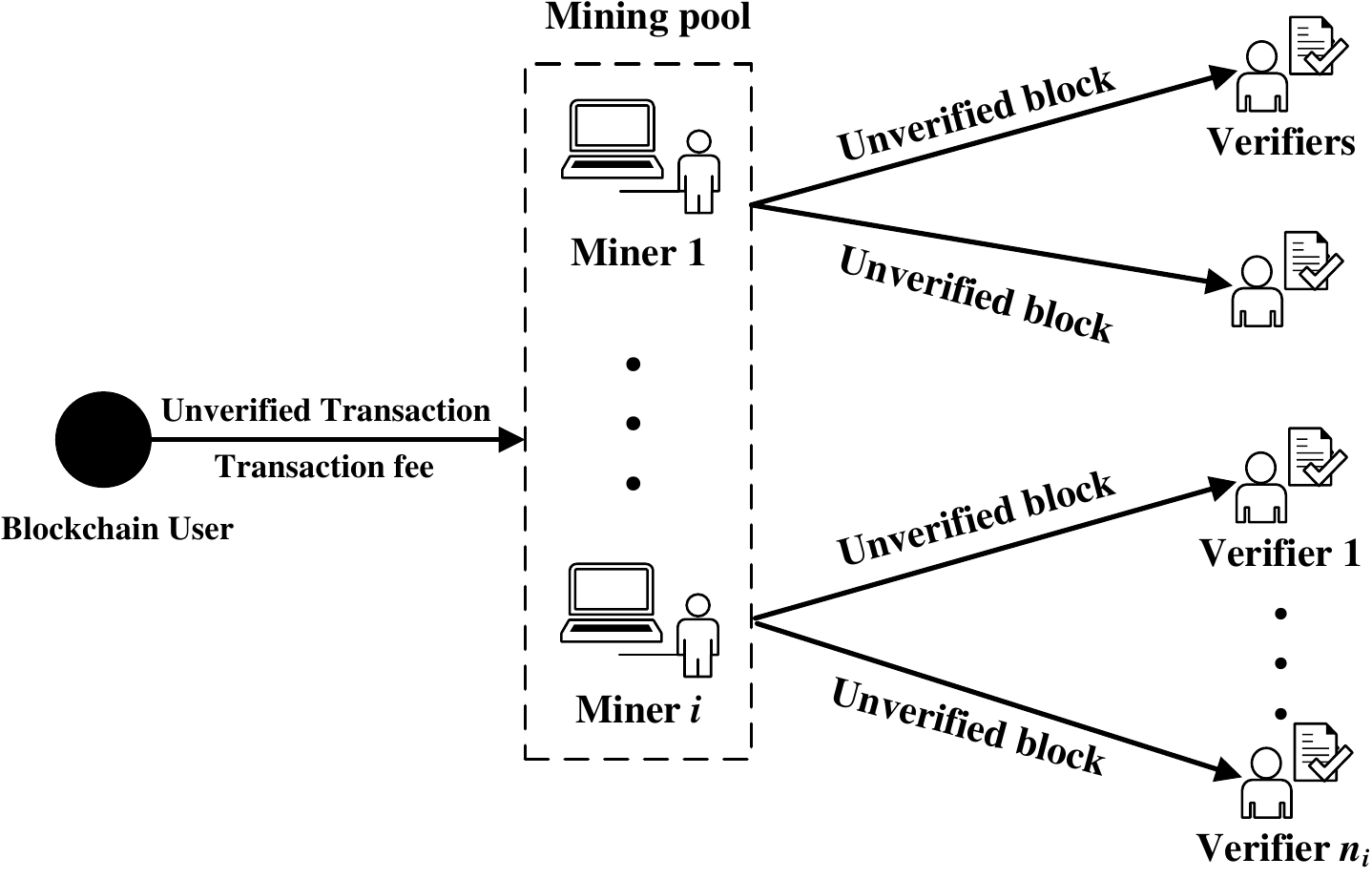}
 \caption{An example that demonstrates the relationship among blockchain user, miners and verifiers in consortium blockchain.  The miners recruit some other miners, i.e.,
verifiers, to verify the transaction \cite{kangwcl}. }
 \label{consortium}
\end{figure}

The majority attack also exists in the PoS based consortium blockchain \cite{zheng2017overview}. In the system, the blockchain user produces transactions for verification and pays the transaction fee. Due to the limited number of miners, some miners can launch the majority attack, i.e., halt or reverse transactions by forming a cartel. Thus, in addition to competing to solve the crypto-puzzle, the pre-selected miners recruit some other miners, i.e., verifiers, to verify the transaction. This results in recruitment cost and propagation delay that reduce the utility of the pre-selected miners \cite{li2017consortium}. In this case, the blockchain user acting as the leader sets the transaction fee for relative secure verification. The pre-selected miners acting as the followers, given the other miners' strategies, trade off the transaction propagation delay and recruitment cost against the transaction fee offered by the blockchain user. This scenario is illustrated in Fig \ref{consortium}. Therefore, the interaction between the blockchain user and the pre-selected miners can be modeled as a Stackelberg game as presented in \cite{kangwcl}. By using the second-order derivation, the blockchain user and pre-selected miners' utility functions are proved to be concave. Thereby, they can jointly maximize their utility through backward induction. The simulation results show the bigger variation range of propagation delay brings lower utility of the blockchain user. However, the game model is under the assumption of complete information of the all miners' strategy. The Bayesian game model \cite{armbruster1979bayesian} can be used to analyze the incomplete information case.

\subsection{Denial of Service (DoS) Attack}
Due to the distributed structure of peer-to-peer (P2P) network in blockchain with the Nakamoto consensus protocol, each miner can observe the PoW done by their peer miners \cite{nakamoto2008bitcoin}. However, if the P2P network is interfered or disrupted by some attackers, the attacked miners' resources available for transaction propagation and verification may be exhausted. Thus, the attacked miners would not complete the mining process to gain the mining rewards and their expected profit. Such an attack is called Denial-of-Service (DoS) \cite{vasek2014empirical}.

The mining pools can perform the DoS attack as presented in \cite{johnson2014game}. More specifically, to maximize the mining reward, the mining pools can choose (i) to trigger the distributed DoS (DDoS) attack that lowers the other mining pools' expected payoff, or (ii) to invest in additional computational power, e.g., by buying more mining machines, to increase its possibility of solving the next PoW. Each mining pool trades off the cost of the investment and attack associated with the other pools' strategies, as well as the uncertainty of launching the attack successfully. Therefore, a non-cooperative game can be adopted to analyze the interaction among the pools with different sizes. In the game, there are two players, i.e., a big pool and a small pool. The other pools own the rest of computational power. The payoff of different strategies of the two players can be expressed in a matrix in terms of the computational power distribution, the increasing rate of network's computational power over time, and the probability of launching the DDoS attack successfully. This matrix is presented in Table \ref{matrix} where $P_s$ and $P_b$ are the payoffs of the small pool and the big pool, respectively.

\begin{table}[!htbp]
\caption{Payoff matrix with launching DDoS attack}\label{matrix}
\centering
\begin{tabular}{|c|c|c|}
\hline
&Investment(I)&Attack(A)\\
\hline
Investment(I)&$P_s(I,I)$,$P_b(I,I)$&$P_s(I,A)$,$P_b(I,A)$\\
\hline
Attack(A)&$P_s(A,I)$,$P_b(A,I)$&$P_s(A,A)$,$P_b(A,A)$\\
\hline
\end{tabular}
\end{table}

Since investing in computational power is the only best response of both big and small pools when computational power distribution satisfies an inequality and vice versa for launching the DDoS attack, the unique Nash equilibrium can be obtained under different computational power distribution. Simulation results show that mining pools have different incentive to perform DDoS attack under different computational power distribution. Due to the higher expected payoff, each pool has a greater incentive to attack larger pools than smaller ones and the larger mining pools have a greater incentive to perform the DDoS attack than smaller ones. These results are consistent with the empirical evidence on the prevalence of DDoS attacks in the Bitcoin system as presented in \cite{vasek2014empirical}. The authors in \cite{johnson2014game} also consider the incentive of mining pools as a whole. However, in a real case, the individual miners have an incentive to hop among the pools and then the computational power distribution changes. Thus, the Nash equilibrium may be shifted.

Apart from only focusing on the short-term impact of DDoS attacks on mining pools as presented in \cite{johnson2014game}, the authors in \cite{laszka2015bitcoin} study the long-term impact. An ongoing DDoS attack causes some long-term impacts that individual miners may migrate, i.e., leave the attacked pool and participate in other pools. The model consists of two pools. At every stage of mining competition, each pool chooses an attack level, i.e., the fraction of its computational power to launch the attack to the other pool. Choosing the attack level affects both the short-term utility consequences (as studied in \cite{johnson2014game}) and the long-term consequences. In particular, the long-term consequences affect the computational power distribution of mining pools in the next stage. Therefore, the interaction between the two pools can be modeled as a sequential game. By using the second-order derivative, the utility function of the mining pool, i.e., the player, is proved to be concave under the condition that the attack cost is greater than a certain value. This value is associated with the level of attracting miners to participate in the pool, and the migration rate of miners that are not affected by the attack. Thus, the game can reach a unique Nash equilibrium at which both the mining pools have no incentive to launch the DDoS attack. However, if the condition is not satisfied, the game reaches another Nash equilibrium at which one of the players attacks while the other remains not attacking. For the future work, a general case of multiple mining pools can be investigated.

To avoid such DDoS attack, the authors in \cite{nojoumian2018incentivizing} propose a reputation-based scheme in which each miner is assigned a reputation value that evaluates the miner's performance of mining honestly against launching DDoS attack. The pool managers send invitation probabilistically only to a subset of miners according to the miners' reputation values. Only miners that receive invitations from pool managers can mine for the pool. Otherwise, the miner has to mine for itself, and this is not preferable for the miner with small computational power. To maximize the profit, each miner chooses to attack or mine honestly while optimizing the profit of launching attack and minimizing the probability to be excluded from pool managers' invitation because of the decrease of its reputation value from the attack. Since the reputation value is updated periodically, and each miner determines its strategy based on the future utility associated with the other miners' reputation value and strategy, a repeated game is used to model the interaction among miners. By removing the strictly dominated strategies of the game according to the miner's utility function, the unique Nash equilibrium can be obtained such that the best response of each miner is not to launch the attack. Similar to the analysis presented in \cite{laszka2015bitcoin}, the reason is that even the miner can gain some utilities in the current stage of mining competition by launching the attack, the miner will lose many future mining opportunities due to its lower probability of being invited to mine for the pool. However, implementing this reputation-based scheme through the simulation-based approach is not discussed in the paper.

Similar to \cite{nojoumian2018incentivizing}, a punishment scheme based on the action record in blockchain to suppress the attack motivation is proposed in \cite{xugame}. Nevertheless, the scheme is applied to an edge network instead of the blockchain system. The network model consists of mobile devices and one server located in the edge network. The mobile devices can (i) send service requests to the server, or (ii) launch the DoS attacks to gain their illegal profits. The server can choose (i) to execute the service requests, or (ii) to launch the attack on the devices. Each device or the server can decide its strategy according to the other's historical strategy recorded in the blockchain. Therefore, the interaction between the mobile device and the server can be modeled as a non-cooperative game. The utility of both the players, i.e., the mobile device and the server, is a function of the cost and profit of launching the attack and executing the request, and a punishment factor related to their historical strategies. Since the players can maximize their utility by not attacking under a certain constraint associated with the punishment factor, not attacking is the best response of the players, and thus the game can reach the Nash equilibrium. Simulation results also show that both mobile device and edge server tend to not attack if the punishment factor is large and the attack rate of the server decreases compared with that of the non-punishment scheme. However, the existence of the Nash equilibrium may not be guaranteed in a multi-player scenario.

\subsection{Other Security Issues}

The underlying blockchain technology of bitcoin is now being applied to many new scenarios such as edge networks, cloud computing, e-business and information sharing \cite{underwood2016blockchain,tschorsch2016bitcoin,crosby2016blockchain}. In particular, a series of security problems regarding false data sharing \cite{rawat2018ishare,dong2017betrayal,adler2018astraea}, distrustful goods trading \cite{asgaonkar2018solving,bigi2015validation} and cyber-insurance \cite{feng2018cyber}, can be resolved by using a blockchain-based scheme.

\subsubsection{False Data sharing}

Blockchain-based scheme is applied to the false data sharing scenario. In most of the traditional data sharing application scenarios, the users transfer data either to other users or to a centralized authority for verification. However, the users are reluctant to share the cyber-security information due to the concern about the distrust, the possible false information, the privacy vulnerabilities, and the lack of incentive \cite{rutkowski2010cybex}. To address these problems, the authors in \cite{rawat2018ishare} propose a blockchain-based information sharing (iShare) framework. In the iShare framework, organizations, i.e., users participating in sharing cyber-attack information, receive a reward after the information transaction is proved authentic in the blockchain. The organizations can form a group to share information and gain the reward together similarly to forming the mining pool in bitcoin systems \cite{rosenfeld2011analysis}. However, some group members can form a sub-group and infiltrating in another group to gain more profit by not releasing the information in the infiltrated group. This is similar to launching the PBWH attack (see Section \ref{SMA}) as presented in \cite{eyal2015miner}. In the two-group case, each group determines the number of organizations to infiltrate to the other group to maximize its profit. Thus, the non-cooperative game can be used to analyze the interaction between the two groups. Each group's utility is determined based on the size and number of infiltrating organizations of the two groups. Since the utility function of the group is concave, each group can maximize its profit when the number of infiltrating organizations satisfies the first-order optimality condition. The unique Nash equilibrium can be obtained at this point in which not launching the attack can be the best response for each group. The Nash equilibrium shifts when the number of infiltrating organizations satisfies the different constraint and thus launching attack can also be the best response for the group. However, a general case of multi-groups can be investigated for the future work.

False information risk among the users and the lack of incentive can also be found in the traditional cloud computing scenario. The cloud users may not entirely trust the computing results returned from the cloud provider. Thus, the verifiability becomes a critical requirement by the cloud users. The existing techniques, e.g., \cite{walfish2015verifying}, for verifying correctness of the result cannot be done at a reasonable cost. A blockchain-based scheme with smart contract can be used to address the issue as proposed in \cite{dong2017betrayal}. In the scheme, the cloud user pays two clouds, using smart contract, for computing the same task and then collects and crosschecks the results from the two clouds to verify the correctness. However, the two clouds can collude with each other, i.e., output the same wrong result, to gain an extra profit. To maximize the utility, each cloud chooses to compute honestly or to collude to trade off the profit obtained from the cloud user's payment and the loss of the deposit, i.e., a sum of money that guarantees the security for the delivery of the correct result. The cloud's expected utility function is determined based on not only its present strategy but also the imperfect information of the other clouds' historical strategies over time. Thus, the extensive-form game can be used to analyze the interaction between the two clouds. By using the backward induction, each cloud is proved to obtain the strictly dominant strategy that maximizes its utility function at every information set in every sub-game, and thus the game can reach the unique sequential equilibrium. At the sequential equilibrium, both clouds have no incentive to deviate from computing honestly, i.e., not to collude. Simulation results show that the proposed scheme can achieve a low cost compared with the techniques from \cite{walfish2015verifying}. The reason is that the cloud users only need to pay the cost of employing two clouds for computing the same task.

Nevertheless, although the smart contract has the advantages as presented in \cite{dong2017betrayal}, a major limitation exists that only data in the blockchain is processed, and trusted entities are required to verify the correctness of the external data that will be brought into the blockchain. The trusted entities can launch an attack by manipulating the data to gain an extra profit \cite{peterson2018augur}. The authors in \cite{adler2018astraea} propose a decentralized entity scheme to prevent the attack. The model consists of the voters and the verifiers. The voter can vote, i.e., state the data as either true or false, for random data once the voter submits a small deposit to the system. The verifier can vote for the chosen data after submitting a large deposit. Each participant, i.e., the voter or the verifier, can receive a reward if its correctness statement is the same as that of the other participants. Thus, a coordination game can be used to analyze the interaction between the voter and the verifier. According to the definition of the coordination game \cite{han2012game}, it can be easily proved that the game has two Nash equilibria in which the participants state the same correctness. At the Nash equilibrium, rational participants have no incentive to deviate from voting honestly if the majority participants give the honest statement. The simulation results show that the proposed game can achieve a zero probability of data manipulation.

\subsubsection{Distrustful Goods Trading}

The distrust of goods trading can also be mitigated by applying blockchain based smart contract as presented in \cite{asgaonkar2018solving}. The proposed smart contract involves two participants, i.e., one seller and one buyer. The participants are required to place a sufficiently large deposit for the reliable transaction which will be returned only after the transaction is completed. The participants can choose to cooperate, i.e., execute the transaction honestly, or to attack, i.e., cheat another participant, e.g., by double spending. To maximize the utility, each participant has to take into account the tradoff between the cost, i.e., the loss of the deposit, and the profit of launching the attack given the other participant's strategy. The seller takes its strategy before the buyer does, and thus an extensive-form game can be formulated. The utility of the player, i.e., the seller or the buyer, is determined based on the deposit, the value of the goods and the price set in the smart contract agreement. By using the backward induction, the game is proved to admit a unique subgame perfect Nash equilibrium at which both players perform the transaction honestly. However, how to implement the proposed smart contract is not discussed.

Using a deposit for buying and selling goods can also be found in \cite{bigi2015validation}. The transaction is insured by the deposit of both participants, i.e., the buyer and the seller. The buyer's strategy profile includes (i) PC: pay and confirm the transaction, (ii) PD: pay and leave the system with denying the transaction, and (iii) $L_b$: leave the system without paying. The seller's strategy profile includes (i) SC: ship the goods and confirm the transaction, (ii) SD: ship the goods and leave the system with denying the transaction, and (iii) $L_s$: leave the system without shipping. Each participant's payoff is determined based on the value of the goods and its deposit given the other participant's strategy. The interaction between the two participants can be modeled as a normal-form game. By using the iterative removal of dominated strategies \cite{han2012game}, the game is proved to have a unique Nash equilibrium if both the participants' deposits are greater than the goods' value. At the Nash equilibrium, the PC and SC strategies are the best response of the buyer and the seller, respectively. Simulation results show that if the deposits of both participants are greater than the value of the goods, the sum of buyer's money and the value of the seller's goods remain unchanged for the whole system. This means that the buyer's money is exchanged into the goods successfully, and the seller's goods is exchanged into the money with no loss. However, in practice, the participant may not be perfectly knowledgeable of the other participant's strategy. More sophisticated game models and tools can be considered.

\subsubsection{Cyber-Insurance}

Different from suppressing the attack motivation as presented in \cite{rawat2018ishare,dong2017betrayal,asgaonkar2018solving,bigi2015validation,adler2018astraea}, the authors in \cite{feng2018cyber} propose a cyber-insurance scheme \cite{pal2014will} to compensate for the losses of the attacked blockchain participants. The model includes multiple blockchain users, one blockchain provider, and one cyber insurer. Each user needs to choose a service offered by the provider and maximize its utility given the other users' service demands. Given the users' demand, the provider's problem is to invest in the computing resource to increase its profit. To alleviate losses of being attacked, the blockchain provider also purchases insurance from the cyber-insurer. The cyber-insurer sets the price of the insurance based on the perceived risk level of the provider. Typically, the provider and the insurer offer the service first, and the user then chooses the service. Thus, the interaction among the users, the provider, and the insurer can be modeled as a Stackelberg game. By exploiting the characteristics of the Jacobian matrix \cite{rosen1965existence} to analyze the utility functions of the players, the game is proved to admit a unique Stackelberg equilibrium. The simulation results show that the provider can maximize its utility at a unique point which is in accordance with the uniqueness analysis. However, in practice, the insurer cannot completely know the risk level of the provider, and the Bayesian game can be adopted.

\section{Applications of Game Theory for Mining Management}

Under the Nakamoto protocol, anyone within the blockchain network is allowed to play the role of the mining competition, transaction dissemination and verification in order to obtain the profit \cite{nakamoto2008bitcoin}. Each miner or mining pool involved manages what strategy it will perform to maximize its payoff given the others' strategies, and game theory can thus be effectively applied. In this section, we will survey the applications of game theory in the mining management including computational power allocation, fork chain selection, block size setting, pool selection and reward allocation.

\begin{table*}
\caption{\label{tab: }A Summary of Game Theoretical Applications for Mining Management.}
\label{Mining_Management_summary_table}
\begin{centering}
\begin{tabular}{|>{\centering\arraybackslash}m{0.3cm}|>{\centering}m{0.5cm}|>{\centering\arraybackslash}m{1.9cm}|>{\centering}m{2cm}|>{\centering}m{2.4cm}|>{\centering}m{3.5cm}|>{\centering}m{2.2cm}|>{\centering}m{2cm}|}
\hline
 \cellcolor{mygray}  \textbf{\noun{}} &  \cellcolor{mygray} \textbf{\noun{Ref.}}&  \cellcolor{mygray} \textbf{\noun{Game Model}}&  \cellcolor{mygray} \textbf{\noun{Player}}&  \cellcolor{mygray} \textbf{\noun{Action}}  &  \cellcolor{mygray} \textbf{\noun{Strategy}} &  \cellcolor{mygray}\textbf{\noun{Payoff}} & \cellcolor{mygray} \textbf{\noun{Solution}} \tabularnewline
\hline
\hline

\parbox[t]{2mm}{\multirow{40}{*}{\rotatebox[origin=c]{90}{ \hspace{-5 cm} Individual mining}}}
&\cite{dimitri2017bitcoin}& Non-cooperative game &Miners& Computational power allocation & Selection of investment in computational power or not & Mining rewards minus cost & Nash equilibrium \tabularnewline \cline{2-8}
&\cite{dhamal2018stochastic}& Stochastic game &Miners & Computational power allocation &Selection between investing and leaving&Mining rewards minus cost & Subgame perfect equilibrium  \tabularnewline \cline{2-8}
&\cite{chiu2018incentive}& Cournot game & Miners & Computational power allocation & Determination of the amount of investment in computational power & Mining rewards minus cost & Nash equilibrium \tabularnewline \cline{2-8}
& \cite{tsabary2018gap}& Non-cooperative game & Miners & Computational power allocation & Selection of proper time to start using the mining machines & Mining rewards minus cost & Nash equilibrium \tabularnewline \cline{2-8}
&\cite{xiong2018optimal}& Stackelberg game & Service provider and miners & Computational power allocation & Determination of service price and service demand & Profit minus cost & Stackelberg equilibrium \tabularnewline \cline{2-8}
&\cite{luong2018optimal}& Auction & Service provider and miners & Computational power allocation & Determination of the bid for service & Profit minus cost & Individual utility \tabularnewline \cline{2-8}
&\cite{jiao2018auction}& Auction & Service provider and miners & Computational power allocation & Determination of the bid for service & Profit minus cost & Social welfare \tabularnewline \cline{2-8}

&\cite{beccuti2017bitcoin}& Sequential game & Miners& Fork chain selection & Selection of reporting mined block and mining on the longest chain or not & Mining rewards & Sequential equilibrium \tabularnewline \cline{2-8}
&\cite{biais2018blockchain}& Stochastic game & Miners & Fork chain selection & Selection of branch to mine & Mining rewards & Subgame perfect equilibrium  \tabularnewline \cline{2-8}
&\cite{saleh2018blockchain}& Extensive-form game & Miners & Fork chain selection & Selection of mining on the fork or not & Mining rewards minus cost & Nash equilibrium \tabularnewline \cline{2-8}
& \cite{azouvi2018betting}& Extensive-form game & Miners & Fork chain selection & Selection between strategically or stubbornly deviating the protocol and following the protocol & Mining rewards minus cost and punishment & $\epsilon$-robust equilibrium \tabularnewline \cline{2-8}
&\cite{barrera2018blockchain}& Coordination game & Miners & Fork chain selection & Determination of updating blockchain version or not & Mining rewards & Nash equilibrium \tabularnewline \cline{2-8}
&\cite{abadi2018blockchain}& Coordination game & Blockchain users and miners & Fork chain selection & Chosen between two fork chains & Mining rewards & Nash equilibrium \tabularnewline \cline{2-8}
&\cite{stone2018delayed}& Repeated game & Miners & Fork chain selection & Selection of forming the coalition or not & Mining rewards minus cost & Social welfare \tabularnewline \cline{2-8}
&\cite{pass2017fruitchains}& Non-cooperative game & Miners & Fork chain selection & Selection of forming the coalition or not & Profit minus cost & $\rho$-coalition-safe 3$\delta$ Nash equilibrium \tabularnewline \cline{2-8}
&\cite{kiayias2017ouroboros}& Non-cooperative game & Miners & Fork chain selection & Selection of forming the coalition or not & Mining rewards & Nash equilibrium \tabularnewline \cline{2-8}

& \cite{houy2014bitcoin}& Non-cooperative game & Miners & Block size setting & Determination of the block size & Transaction fees and mining rewards & Nash equilibrium \tabularnewline \cline{2-8}
&\cite{houy2014economics}& Non-cooperative game & Miners & Block size setting & Chosen transaction to be included in the block & Transaction fees and mining rewards & Nash equilibrium \tabularnewline \cline{2-8}
&\cite{zhang2017necessity}& Non-cooperative game & Miners & Block size setting & Chosen of upper bound of block size & Transaction fees & Nash equilibrium \tabularnewline \cline{2-8}
&\cite{lewenberg2015inclusive}& Extensive-form game & Miners & Block size setting & Chosen transaction to be included in the block & Transaction fees & Sequential equilibrium \tabularnewline \cline{2-8}
&\cite{easley2017mining}& Non-cooperative game & Blockchain users & Block size setting & Selection of paying the transaction fee or not & Profit minus transaction fee & Nash equilibrium \tabularnewline \cline{2-8}
&\cite{abraham2016solidus}& Non-cooperative game & Miners & Block size setting & Selection to be included in the committee & Mining rewards & Nash equilibrium \tabularnewline \cline{2-8}
\hline

\parbox[t]{2mm}{\multirow{5}{*}{\rotatebox[origin=c]{90}{ \hspace{-3 cm} Pool mining}}}
&\cite{lewenberg2015bitcoin}& Coalitional game & Miners and pools & Pool selection & Chosen of the pool to join & Mining rewards & Cooperative equilibrium \tabularnewline \cline{2-8}
&\cite{liu2018evolutionary}& Evolutionary game & Miners and pools & Pool selection & Chosen of the pool to switch & Mining rewards minus cost & Nash equilibrium \tabularnewline \cline{2-8}
&\cite{brunjes2018reward}& Coalitional game & Miners and pools & Pool selection & Selection between forming the pool and joining the pool&  Mining rewards & Non-myopic Nash equilibrium\tabularnewline \cline{2-8}
&\cite{schrijvers2016incentive}& Non-cooperative game & Miners and pool manager & Reward allocation & Selection of reporting shares and allocating rewards & Mining rewards & Nash equilibrium \tabularnewline \cline{2-8}
&\cite{fisch2017socially} & Repeated game & Miners and pool manager & Reward allocation & Selection of reporting shares and allocating rewards & Mining rewards & Nash equilibrium\tabularnewline \cline{2-8}
&\cite{cong2018decentralized}& Extensive-form game & Miners and pool manager & Reward allocation & Determination of the computational power allocation and optimizing the reward allocation & Mining rewards minus cost and charged fee & Subgame perfect equilibrium \tabularnewline \cline{2-8}

\hline

\end{tabular}
\par\end{centering}
\end{table*}

\subsection{Individual Mining}

\subsubsection{Computational Power Allocation}

Bitcoin mining is a competition that miners contend with each other by investing in computational power to win mining rewards. To maximize the utility, each miner determines the allocation of its computational power, i.e., whether or not to invest in the computational power, given the other miners' strategies. Therefore, a non-cooperative game is applied to analyze the interaction among the miners in \cite{dimitri2017bitcoin}. The miner's utility is a function of its computational power, the mining rewards and the marginal cost, i.e., the average cost for the miner to invest in a unit of computational power. By using the second-order derivative, the miner's utility function is proved to be concave. Thus, a unique Nash equilibrium exists at which investing is the best response of each miner as long as the miner's computational power satisfies a condition. Here, the condition is determined based on the computational power and the marginal cost of the miner and the entire bitcoin network. At the equilibrium, it is found that the decision on the investment is not affected by the value of the mining rewards. Moreover, every miner can have a positive utility for any level of other miners' strategies which consequently can prevent a monopoly.

Different from \cite{dimitri2017bitcoin} in which the miners choose whether or not to participate and then keep their chosen strategies, the authors in \cite{dhamal2018stochastic} consider a case in which the miners can choose ``arrival'', i.e., investing in the computational power, and ``departure'', i.e., leaving the mining, at any time. In general, the strategy of each miner depends on the state of the blockchain network, i.e., the number of miners participating in the mining, given other miners' strategies. A stochastic game can be applied to analyze the miners' strategies as presented in \cite{dhamal2018stochastic}. The miner's utility is a function of the number of the miners in the system, the arrival and departure rates of the miners, the rate of PoW getting solved, the cost and the reward of the mining. By transforming the utility function to the Bellman equation \cite{bellman1952theory} and then calculating the first-order derivative, the utility function is proved to be monotonic increasing if the cost of mining is greater than a threshold. Thus, investing the maximum power is the dominant strategy of each miner regardless of the state of the blockchain network, and the game has a subgame perfect equilibrium. The simulation results show that the utilities of the miner under different arrival rates gradually converge to the same curve, i.e., the game reaches the equilibrium.

In addition to the case in \cite{dimitri2017bitcoin} and \cite{dhamal2018stochastic} that the miner can only choose to invest in the computational power or not, the authors in \cite{chiu2018incentive} investigate the amount of computational power that the miner determines to invest to win the mining rewards, given the other miners¡¯ strategies. The probability that the miner solves the PoW in a given time can be assumed to follow an exponential distribution \cite{nakamoto2008bitcoin}. As such, the Nakamoto protocol essentially formalizes an exponential race. A Cournot game \cite{allaz1993cournot} can be thus used to analyze the interaction among the miners as presented in \cite{chiu2018incentive}. The miner's utility is a function of the mining rewards, the computational power, and the marginal cost of the investment. The game is then proved to admit a symmetric Nash equilibrium by simply showing that the marginal revenue, i.e., the average revenue for the miner to invest in a unit of computational power, is equal to the marginal cost. At the equilibrium, each miner can optimize its investment and has no incentive to deviate from honest mining.

The aforementioned approaches, i.e., \cite{dimitri2017bitcoin}, \cite{dhamal2018stochastic} and \cite{chiu2018incentive}, consider the case that the mining reward dominates the transaction fee. Nevertheless, when the transaction fee dominates the mining reward\footnote{Take the bitcoin as example, the bitcoin code includes a statement which declares that the mining reward will drop by half after about four years (210,000 blocks). Thereby, the mining reward will eventually be dominated by the transaction fee.}, the miner will adjust its allocation of computational power by choosing strategically the proper time to start using its mining machines, i.e., the machines used for mining process which require electricity for their operation, to mine given the other miners' strategies. The reason is that miners have no incentive to mine unless the accumulated transaction fees sufficiently exceed a certain threshold. Thus, the non-cooperative game can be used to analyze the interaction among the miners as presented in \cite{tsabary2018gap}. Each miner's utility is a function of the starting time, the operation time, the proportion of the miner's machines, and the probability distribution function of the block finding time. The numerical analysis is thus used to find the Nash equilibrium of the game. The simulation results show that the miners that own the same number of mining machines eventually converge to the same starting time, meaning that the game reaches the Nash equilibrium. However, how to prove the uniqueness of the Nash equilibrium is not discussed.

\begin{figure}[htbp]
 \centering
\includegraphics[width = 5.8 cm, height = 4 cm]{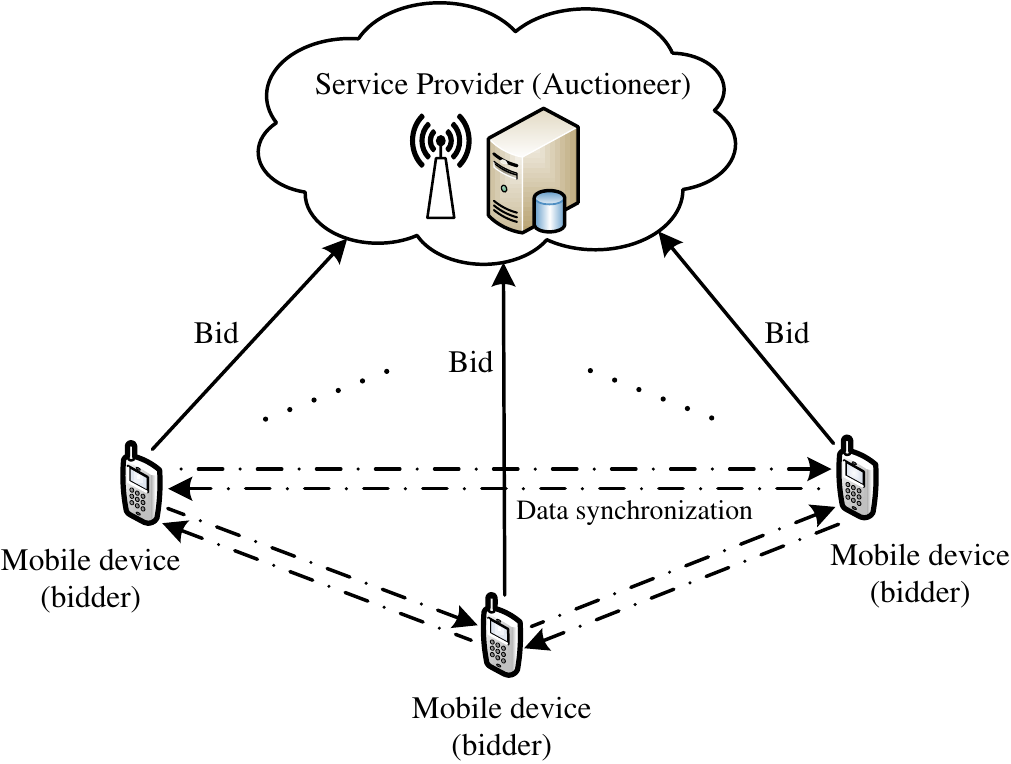}
 \caption{ An example of the system model of edge computing in mobile blockchain network. The mobile devices compete for the computational power by submitting the bid, and the service provider determines the allocation rule of its service.}
 \label{ET}
\end{figure}

Although blockchain has been widely deployed in many scenarios as presented in \cite{dimitri2017bitcoin,dhamal2018stochastic,chiu2018incentive,tsabary2018gap}, deploying blockchain applications in mobile environments is still challenging because the mining process consumes high computational power from mobile devices. An edge computing paradigm has been recently introduced in the mobile blockchain networks for offloading the mining tasks of mobile devices, i.e., the miners \cite{xiong2018mobile}. The system model is illustrated in Fig \ref{ET}. However, an important issue is how to allocate efficiently the limited edge computing resources of service providers to the miners. The authors in \cite{xiong2018optimal} model the interaction among the service provider and the miner as a two-stage Stackelberg game. The service provider acts as the leader setting the price of the service, and then the miner acts as the follower choosing its computational service demand, given the service price and the other miners' strategies. The utility of service provider is a function of the profit obtained from charging the miners, the miners' service demand, the time that the miner takes to mine a block, and the cost of electricity. The utility of the miner is a function of the computational service demand, the service price, the cost and the rewards of the mining. By using the backward induction, the game is proved to admit a unique Stackelberg equilibrium which is supported by the simulation results. However, in practice, the players cannot know the perfect information of each other, and the Bayesian game can be adopted.

Traditional sealed-bid auctions, e.g., the Vickrey auction \cite{vickrey1961counterspeculation}, can also be used to guarantee that the edge computing resources are allocated to the miners which value the resources most. However, designing the optimal auction is challenging. The authors in \cite{luong2018optimal} propose to apply deep learning techniques to achieve the optimal auction for the computing resource allocation in the blockchain network. The model consists of one service provider, i.e., the seller or auctioneer, and multiple mobile users as miners, i.e., bidders. The miners compete a computing resource unit of the service provider by submitting bids, i.e., the prices that the miners are willing to pay. Upon receiving the bids, the service provider determines the allocation rule, i.e., winning probabilities of the miners, and the conditional payment rule to the miners. The allocation and payment rules are implemented by using neural networks. The neural networks are constructed based on an analytical solution of the optimal auction, i.e., the Myerson theory \cite{myerson1981optimal}. As such, the auction mechanism learned by the neural networks is optimal in terms of maximizing the revenue of the service provider while ensuring the economic properties, i.e., incentive compatibility and individual rationality. The simulation results show that the proposed scheme outperforms the traditional sealed-bid auction \cite{vickrey1961counterspeculation} in terms of revenue. However, the proposed scheme is constrained to a single computing resource unit that may not meet the needs of the miners.

Different from the auction in \cite{luong2018optimal} that the service provider, i.e., the auctioneer, maximizes its individual utility, the authors in \cite{jiao2018auction} investigate the case of maximizing the social welfare of the entire blockchain network. Under the same model as that in \cite{luong2018optimal}, the utility of the mobile user and service provider is a function of the mining rewards, the computational power, the service price, the demand of the miner, and the robustness of the network associated with the distribution of the computational power. By transforming the social welfare maximization auction problem to a problem of non-monotone submodular maximization with knapsack constraints \cite{lee2009non}, the algorithm of achieving the social optimum can be developed. The simulation results show that the algorithm not only achieves the good performance in maximizing the social welfare, but also guarantees the truthfulness, individual rationality and computational efficiency. However, the algorithm is designed for the offline auction which is not applicable for real-time trading scenarios.

\subsubsection{Fork Chain Selection}

Under the Nakamoto protocol, there are sequential PoW puzzles that the next puzzle depends on the solution of the previous one. Since each miner needs to choose to (i) report its found puzzle to mine on the longest chain, or (ii) not to report the found puzzle and to the next puzzle secretly, given the publicity of previous puzzles, fork chain may appear. To maximize the utility, the miner trades off reporting the puzzle to gain the mining rewards and not reporting the puzzle to mine on fork. Meanwhile, the miner is uncertain whether it is the first one to find the solution of the puzzle. Thus, a sequential game with imperfect information can be applied to model the interaction among the miners as presented in \cite{beccuti2017bitcoin}. The miner's utility is a function of the distribution of the computational power, the probability of winning to solve the PoW, and the other miners' belief of the upcoming publicity of the puzzles. By using the backward induction, the game is proved to admit a multiplicity of sequential equilibrium. This means that both reporting and not reporting can be the best response of each miner depending on the computational power that the miner uses to solve the puzzle. However, the authors only consider a three-miner case, and a general case with any number of miners can be investigated.

After finding the solution of the PoW as discussed in \cite{beccuti2017bitcoin}, the miner probabilistically chooses which branch to mine, i.e., to choose a certain chain to attach its block to, among the tree-like branches of the blockchain network structure. If the miner chooses the branch which will not be the longest chain, the miner's effort to solve PoW is wasted. A stochastic game can be used to analyze the strategies of the miners as presented in \cite{biais2018blockchain}. The miner's utility is determined based on the miner's computational power, the number of blocks solved by the miner, the mining rewards, and the difficulty of solving the PoW. By using the backward induction, the game is proved that mining the longest chain is a subgame perfect equilibrium. However, the current longest chain may not be the longest one after several rounds of mining competition. Portions of the historical transactions may be abandoned.

Similar to \cite{biais2018blockchain}, the authors in \cite{saleh2018blockchain} demonstrate that following the protocol, i.e., mining on the longest chain, is the Nash equilibrium. However, the model in \cite{saleh2018blockchain} is based on the PoS system in which the fork chain randomly selects the coin from the set of coins owned by miners at each time step (see Section \ref{background}). Thus, an extensive-form game can be applied. The miner chooses whether or not to mine on the fork, given the other miners' strategies. The miner's utility is a function of the stake, the mining rewards, the coins of miners selected by the fork, and a discounted factor. Since the cost of mining on the fork increases with the miner's stake, for a sufficiently large stake of the miner, the cost overweighs the profit gained from the mining rewards. By restricting access to the miners with the large stake, the rest of miners have no incentive to deviate from mining on the longest chain, and the game thus reaches the Nash equilibrium. Empirical data obtained from Blockchain.info \cite{blockchaininfo} supports the theoretical analysis.

Extended from \cite{saleh2018blockchain}, the authors in \cite{azouvi2018betting} investigate the case of miners choosing the fork chain in an upgraded PoS system. In the upgraded system, the latest block is called the parent block, and concurrent blocks attached to the parent block are called the leaf blocks. Instead of following the longest chain protocol, miners can choose the leaf blocks to be attached to the parent block. To model the interaction among the miners in the tree-like structure of the system, an extensive-form game can be applied. The miners' strategies include deviating from the protocol stubbornly, following the protocol, and strategically choosing whether or not to deviate from the protocol to maximize their utility. Since there is only one leaf block that can reach the consensus to win the reward, the utility of the miner is a function of the reward, the cost of losing the block, and the punishment of deviating from the protocol, given the other miners' strategies. The punishment is implemented by taking away the deposit of the miner that is deposited in advance. When the fraction of the stubborn miners is less than $1/3$, each miner cannot increase its utility more than $\epsilon$ or decrease its utility more than $1/\epsilon$ by deviating from the protocol. Thus, the game has a unique $\epsilon$-robust equilibrium \cite{abraham2016solidus}. The simulation results show that only when the fraction of the deviated miners is greater than a quarter, the utilities of the miners that follow the protocol decrease, as the number of the deviated miners increases. This is consistent with the theoretical analysis.

Furthermore, when the fork chain appears, the miners need to decide whether to update the blockchain version, i.e., to acknowledge the fork as a hard fork or not. The hard fork is a permanent divergence from the previous version of the blockchain which requires the miners to upgrade the blockchain software. Since having more miners participating in a particular chain version increases the value of the version, the miner's strategy depends on not only its individual profit, but also the other miners' profits. Thus, a coordination game can be used as presented in \cite{barrera2018blockchain}. In the game, the miner gains a zero-profit if the miner's strategy is not consistent with those of the majority of miners, and thus the game is admitted a unique Nash equilibrium. At the equilibrium, every miner chooses to stay on the current version or to upgrade the version. However, organizing the voting of upgrading the blockchain version remains a topic for further research.

Similar to \cite{barrera2018blockchain}, the authors in \cite{abadi2018blockchain} propose to use the coordination game approach for choosing the fork chain. However, the players in the game in \cite{abadi2018blockchain} include blockchain users and miners. To maximize the utility, both types of players need to choose between two fork chains to participate. Here, the utility of a blockchain user is a function of the users' distribution of choosing certain chain, the transaction fees, and the strategies of the miners. The miner's utility is a function of the distribution of the users between two fork chains, the computational power, the mining rewards, and the other miners' choice of the chain. If the number of the blockchain users choosing a certain chain is greater than a threshold, the utility of the players can be proved to be monotonous. Thus, the game has a unique Nash equilibrium that all of the players choose the same chain. Otherwise, a mixed strategy Nash equilibrium exists such that players choose the chain randomly. The simulation results show that the user will choose to remain on a certain chain when the number of the users on this chain is greater than a certain value which is in accordance with the theoretical analysis. However, the case which involves multiple fork chains can also be investigated.

The aforementioned approaches, i.e., \cite{barrera2018blockchain} and  \cite{abadi2018blockchain}, show that the miners can coordinate, i.e., through forming a coalition, to increase their utilities by deviating from the honest mining. To address this issue, the authors in \cite{stone2018delayed} propose an upgrade scheme for the blockchain protocol. In the upgrade scheme, the mining reward is delayed to allocate to the miner that finds the solution of the PoW. Also, the miner can receive variable discounted rewards during several rounds of mining after the miner finds the solution. Extended from the coordination game model in \cite{abadi2018blockchain} to its infinite form, a repeated game is then adopted in \cite{stone2018delayed} where each miner chooses whether to form the coalition or not in every round of mining. The utility of each miner is a function of its computational power, the mining rewards, the difficulty of solving the PoW, the cost of mining, the number of rounds for allocating the discounted rewards, and the discounted factor of the rewards. It is approved that if the discounted factor meets an inequality, the miner's utility of honest mining is greater than that of forming the coalition. This means that the game has a unique subgame perfect equilibrium at which the inequality is satisfied, and all the miners perform honest mining.

Similar to \cite{stone2018delayed}, the authors in \cite{pass2017fruitchains} propose a scheme to prevent the miners from forming the coalition. In the scheme, the transactions are first included in a buffer block, and the miner mines on the buffer block by solving the PoW. Only after the buffer block is broadcast and verified, this buffer block becomes the real block and will be attached to the blockchain. The miner can choose whether or not to form the coalition, i.e., deviating from the honest mining, given the other miners' strategies. Thus, the interaction among the miners can be modeled as a non-cooperative game. The miner's utility is a function of the computational power, the number of the blocks in a round of mining, the difficulty of solving the PoW, the distance between the buffer block and the blockchain, the cost and the rewards of mining, and the transaction fees. By calculating the ratio of the upper bound to the lower bound of the coalition's profit, the multiplicative increase in utility is proved to be less than $(1+3\delta)$. Here, the coalition controls less than a $\rho<1/2$ fraction of the computational power, and the constant satisfies $\delta<0.3$. This means that no coalition that controls less than a fraction $\rho$ of the computational power can gain more than a factor $(1+3\delta)$ of the mining rewards and transaction fees by deviating from the protocol. Therefore, the game has a $\rho$-coalition-safe 3$\delta$ Nash equilibrium.

Different from the PoW based coalition as discussed in \cite{barrera2018blockchain,abadi2018blockchain,stone2018delayed}, the coalition in the PoS based system is investigated in \cite{kiayias2017ouroboros}. In PoS, the miner's stake, i.e., a parameter associated with the amount of the miner's cryptocurrency and the time that miner has been holding the cryptocurrency, is updated at the end of each round of mining and the stake will be reset to zero after the miner discovers the block (see Section \ref{background}). The higher stake means less difficulty in mining the block. Thereby, the miner chooses whether or not to form the coalition for holding more stakes to lower the mining difficulty, given the other miners' strategy. Thus, a non-cooperative game can be applied. The miner's utility is a function of the stake, the mining rewards, the number of times that the miner discovers the block and transactions to be included in the block. Since the miners of the coalition, even deviating from the protocol, cannot obtain the utility which exceeds that of the non-coalition, the game is thus proved to have a unique Nash equilibrium at which every miner follows the protocol. However, forming the coalition is not the only way to increase the miner's stake. To increase the holding time and thereby increase the stake, the miner has an incentive to hold its cryptocurrency without mining. As a result, there is no mining miner, and the entire blockchain network crashes.

\subsubsection{Block Size Setting}

When mining the bitcoin, the miner can earn more transaction fees by including more transactions in its block. However, it also decreases the miner's probability of gaining the mining reward \cite{rizun2015transaction} for a number of reasons, e.g., resulting in a longer propagation time for reaching a consensus. Each miner needs to determine strategically the block size, i.e., the number of transactions to be included in a block, to maximize its utility, given the other miners' strategies. Thus, the authors in \cite{houy2014bitcoin} model a two-miner case as a non-cooperative game. The miner's utility is a function of its computational power, block size, and the time to reach the consensus. Since the first-order derivative of the miner's utility function with respect to the block size is always less than zero when the unit transaction fee and the mining reward meet a certain condition, the strategy that all of the miners include no transaction in their block is a unique Nash equilibrium. However, if the transaction fee or the mining reward change, the Nash equilibrium shifts to the strategy that all of the miners include transactions in their block.

To avoid the case that all miners include no transaction in their block as presented in \cite{houy2014bitcoin}, the authors in \cite{houy2014economics} demonstrate the necessity of setting the maximum block size. Same as the game approach as presented in \cite{houy2014bitcoin}, the miner chooses the transactions to be included in a block at every round of the mining competition. The miner's utility is a function of its computational power and the transaction fees associated with the block size and the bitcoin mining reward. The transactions that one miner does not include in its block will be included by another miner before the next round of the mining competition. Thus, when the block size is unlimited, the strategy of including all transactions by all the miners regardless of the fee is the unique Nash equilibrium. It is also found that unbounded transaction fee leads to the same Nash equilibrium. However, inflations of the computational power distribution may have an impact on the existence of the Nash equilibrium of the game.

An analysis of setting a proper block size can also be found in \cite{zhang2017necessity}. The authors propose a Bitcoin-unlimited scheme to increase the throughput of the bitcoin system. In the scheme, each miner chooses its own upper bound of the block size, and invalidates and discards the excessive block, i.e., the block with the size larger than its upper bound. To maximize the utility, the miner trades off the transaction fees and the probability of its block being orphaned based on its mining power, given the other miners' strategies. Thus, a non-cooperative game can be used to model the interaction among the miners. Since any miner that chooses different upper bound gains zero utility, the game is proved to admit a unique Nash equilibrium at which all miners choose the same upper bound. Since only the blocks with appropriate sizes would be added to the blockchain, the block size under the proposed scheme gradually increases to the maximum limit associated with the network capacity. This means that the divergence on the block size is always bounded and the throughput of the system increases. The simulation results show that if all miners have different bounds, the miners that possess large computational power intend to form a coalition to gain extra profit. However, this is harmful to maintaining the bitcoin's decentralized structure.

However, the unlimited block size \cite{zhang2017necessity} may not lead to a higher throughput of the bitcoin system. The reason is that any two blocks may have collisions, i.e., the miners simultaneously choose the same subset of transactions to be included in the blocks. This situation wastes the computational power for verification and lowers the throughput of the system. To address this issue, the authors in \cite{lewenberg2015inclusive} propose an alternative bitcoin protocol. In the protocol, the system selectively incorporates transactions of off-chain blocks into the main chain and awards creators, i.e., miners, of the accepted transactions even if the creators' blocks are not part of the main chain. Each miner chooses the transactions to be included in its block and trades off the transaction fees and probability of the collision. The miners are partially aware of other miners' strategies and take their strategies sequentially. Thus, an extensive-form game can be used to model the interaction among the miners. The utility of the miner is a function of the position of its block in the main chain, the discount factor, and the fees of the chosen transactions. By using the backward induction, the game is proved to admit a sequential equilibrium at which the miners probabilistically choose the transaction to minimize the collision. As a result, the proposed protocol achieves a higher throughput which is consistent with the simulation analysis. However, the game has several other Nash equilibria at which the miners' utilities are much less than that of the sequential equilibrium.

Moreover, even with the unlimited block size as presented in \cite{zhang2017necessity}, there is still a limitation on transactions to be included in the block. The limitation is imposed by the waiting time, i.e., the time that a transaction of the blockchain user waits in a queue to be included in the block. The blockchain user can choose (i) to pay a transaction fee to the miner to reduce the waiting time, or (ii) not to pay any fee and may experience a longer waiting time. The miner can decide to stay on or to leave the mining according to the expected profit of the transaction fees and the cost. Thus, the interaction between the miners and the users can be modeled as a non-cooperative game as presented in \cite{easley2017mining}. The miner's utility is a function of the number of miners in the network, the rate of solving the PoW, the exchange rate between the bitcoin value and the dollar, the transaction fees, the rewards and the cost of the mining. The user's utility is a function of the exchange rate, the transaction fee, the waiting time, the profit of the included transaction, and the fraction of users that pay the fee. The constraint between the number of miners and the rate of solving the PoW can be obtained, when the miner's and the user's utility are both greater than zero. This means that if the constraint is satisfied, the game has a unique Nash equilibrium. At the equilibrium, the miner chooses to stay on the mining and the user chooses to pay the transaction fee. Empirical evidence from blockchain.info \cite{blockchaininfo} is implemented to validate the theoretical analysis. However, multiple Nash equilibria can exist if the constraint is not satisfied.

As presented in \cite{easley2017mining} that the waiting time limits the throughput of the blockchain network, the authors in \cite{abraham2016solidus} propose a novel protocol that greatly reduces the waiting time of the transaction to reach the Nakamoto consensus. In the protocol, there is a committee including a certain number of members, i.e., miners. The block found by any miner is verified only when the majority of members reach the consensus. This miner is then selected as a member in the committee and ranked based on its computational power. The utility of the member is a function of the computational power, the mining rewards and the other members' strategy. Thus, a non-cooperative game can be applied. Since the member gains the positive profit only when the member follows the protocol, i.e., chooses the block with higher rank, the game is proved to admit a unique Nash equilibrium. At the equilibrium, the chain is never forked and the confirmation time for preventing from the double spending is unnecessary. As a result, the throughput of entire the blockchain network increases.

\subsection{Pool Mining}

\subsubsection{Pool Selection}

To reduce the volatility of the mining rewards and to maximize the utility, miners can form a coalition, i.e., mining pool \cite{rosenfeld2011analysis}, and cooperate with the members, i.e., miners in the pool, by following the reward allocation of the pool. Thus, a coalitional game \cite{myerson2013game} can be used to analyze the interaction among the miners and the pools as presented in \cite{lewenberg2015bitcoin}. Since the communication delay of the bitcoin network leads to the non-linearity of the pool's mining rewards, the rewards cannot be distributed stably among the members. This means that there are always some miners having an incentive to leave their pools and join other pools to increase their utility. As a result, no cooperative equilibrium exists in the game. Additionally, as more transactions are processed in the bitcoin system, the non-linearity effect on the mining rewards increases, and thus miners are more likely to switch pools, i.e., select and join the pool which benefits them most.

During pool selection, each miner first randomly selects a mining pool to start mining with and then switches to another pool after a time period according to its expected utility. The distribution of the miners in mining pools of the whole blockchain network evolves over time based on the miners' strategies. Thus, the framework of evolutionary game \cite{weibull1997evolutionary} can be used to analyze the dynamic process of the miners' pool selection as proposed in \cite{liu2018evolutionary}. If the replicator dynamics \cite{hofbauer2003evolutionary}, i.e., the growth rate of the size of the pools, is equal to zero, the distribution of the miners reaches evolutionary stability \cite{hofbauer2009stable}. Here, the utility of the miner is a function of its computational power, propagation delay, the mining reward and the mining cost. By exploiting the characteristics of the Jacobian matrix of the replicator dynamics in a two-mining-pool network, the game is proved to admit conditionally a unique Nash equilibrium, i.e., the evolutionary stability.

The miners in a PoS system can also form coalitions, i.e., pools, to increase their utilities. The miners need to trade off the cost and the expected profit of forming the pool. For this, each miner chooses (i) to form a pool as a leader, or (ii) to allocate its stake to pools that are already created by the other miners given the reward scheme of the system. In particular, the miner first determines the amount of stake to be allocated to be the leader and then calculates the best possible allocation of mining rewards. Thus, a coalitional game can be applied to analyze the respective aspect of interactions among the miners and the pools as presented in \cite{brunjes2018reward}. The results of backward induction illustrate that both the games have a unique non-myopic Nash equilibrium \cite{brams1981nonmyopic}. At the equilibrium, the certain number of pools are formed with the same size. The rewards are distributed evenly among all miners, except for pool leaders that get an additional gain. The simulation results show that starting from no pool, the game quickly converges to multiple pools of an equal size which is consistent with the theoretical analysis.

\subsubsection{Reward Allocation}

Admittedly, the mining pool's reward allocation, i.e., the algorithm used to share mining rewards among miners, has a significant impact on the utilities of the miners \cite{rosenfeld2011analysis}. The miner can choose to report shares, i.e., preimage solutions for a block that meets the requirement set by the pool manager \cite{wang2018survey}, immediately or to delay the reporting given the reward allocation of the pool. The pool manager needs to select the reward allocation algorithm according to the miners' expected utility. Thus, a non-cooperative game can be used to analyze the interactions between the miners and the pool manager as presented in \cite{schrijvers2016incentive}. If a certain condition is satisfied, the strategy that each miner reports the shares immediately is the Nash equilibrium. Here, the condition is associated with the miner's computational power, the probability of finding the full solution of the PoW, the number of reported shares, and the number of the completed rounds of the mining competition.

However, the approach proposed in \cite{schrijvers2016incentive} considers only the single share. Namely, each miner reports the share only one time during mining. In practice, the miners can report the shares repeatedly, and the pool manager can optimize its reward allocation to maximize its utility. Thus, a repeated game can be applied as presented in \cite{fisch2017socially}. The game is proved that the pool manager can use the geometric-pay, i.e., a certain reward function, to achieve the social optimum. The simulation results show that the expected utility of the geometric-pay pool, i.e., the pool that allocates its mining rewards following geometric distribution, is greater than those of both the proportional pay pool, i.e., the pool that shares mining rewards evenly among the shares, and the PPLNS pool which is in accordance with the theoretical analysis.

As the miners participate in mining pools to reduce the volatility of the mining rewards, a large pool may become even larger. It may lead to a centralization against to the fundamental decentralized structure of the blockchain. However, the authors in \cite{cong2018decentralized} demonstrate that this situation will not happen. During each round of mining, the miner chooses to allocate its computational power to a certain pool according to the state of the blockchain, i.e., the distribution of computational power among the pools. The pool manager adjusts the fees charged to the participated miners to maximize its profit, given the state of the blockchain. Thus, an extensive-form game can be applied to analyze the interaction between the miners and the pool managers as presented in \cite{cong2018decentralized}. The miner's utility is a function of the computational power, fee charged by the pool, the distribution of miners among the pools, the cost and the rewards of the mining. If the fee charged by the pool manager satisfies a condition, the game reaches a subgame perfect equilibrium. Here, the condition is associated with the number of the remaining miners in the same pool. At the equilibrium, the large pools charge a higher fee than the small pools. The miners thus choose the small pools to participate to maximize their utility. As a result, the centralization will not happen. Empirical evidence from Bitcoinity and Bitcoin Wiki supports the theoretical analysis.

\section{Applications of Game Theory atop Blockchain platform}

\subsection{Crypto-currency Economic}

\subsubsection{Transaction Transparency}

\begin{figure}[htbp]
 \centering
\includegraphics[width=8 cm, height=3.5cm]{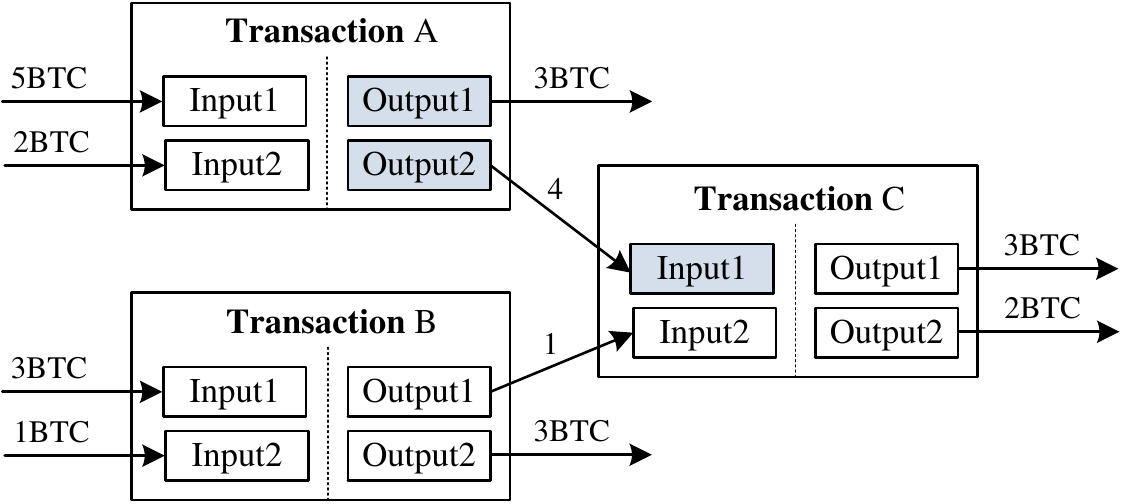}
 \caption{ An example of the mixing payment: the transaction A is discovered to be a ransom payment and all of its outputs are added to the blacklist. The transaction B is in high quality. To avoid transaction C to be orphaned by the miners, the user mixes the payment of the transaction C by the payment of transaction A and transaction B.}
 \label{MT}
\end{figure}

Under the Nakamoto protocol, the entire history records linked to the transaction are transparent to all the blockchain miners and users. This may cause a series of problems. For example, blockchain miners intend to include the transaction in high quality, i.e., most of the history of the transaction is legalized and reliable, into the block rather than the transaction in low quality. The reason is that the transactions that can be traced back to the darknet markets or ransomware payment may be added to the blacklist of the government. The large transaction in not high quality may thus be orphaned by the miners regarding the possible huge loss. To mitigate the risk of transaction in not high quality being orphaned, the user mixes strategically its payment, i.e., splits its payment of transaction into several small ones in different qualities. This scenario is illustrated as in Fig. \ref{MT}. Since the miner's possible loss decreases due to the smaller size of the transaction, the transaction that is not in high quality may be included into the block. The user checks the quality of the other user's transaction sequentially and a sequential game can thus be used to analyze the interaction among the users as presented in \cite{abramova2017mixing}. The user's utility is a function of the quality of the transaction, the value of the post-transaction and the cost of mixing the payment. By using the backward induction, the game is proved to admit multiple subgame perfect Nash equilibria. At the equilibrium, each user mixes their payment in a single transaction instead of sending multiple individual transactions. However, the transaction size, the cost and the rewards of mining can be taken into account in a more general case.

Under the scenario in \cite{abramova2017mixing}, the user that mixes the payment of transaction makes money flows more difficult to trace. This is harmful for the entire blockchain system. To address this issue, the authors in \cite{malinova2017market} investigate the optimal level of transaction transparency and propose a reliable trading system. Since each blockchain user has a unique public key, the user can use the crypto-currency to trade goods with another user directly. To avoid the transaction information, e.g., the ownership of a certain sum of money, being exploited for crime, the proposed system restricts the user's ability to view the complete transaction information attached to the public keys. Thereby, before delivering the goods for trading to other users, the user trades off the expected profit and possible loss in terms of the incomplete transaction information to choose whether or not to perform the trading. The trading can thus be organized as an infinitely repeated game in discrete time as presented in \cite{malinova2017market}. The user's utility is a function of the trading quantity, trading price, the probability of the trade being performed, the allocation of the goods for trading, and the cost of trading. By defining the inequality that the user's utility of offering a positive trading price is greater than that of the offering a zero trading price, i.e., the transaction failure, the constraint between the trading price and the allocation can be obtained. This means that if the constraint is satisfied, the game has multiple Nash equilibria. In any equilibrium of the game, the user has an incentive to split large transaction into small ones and trades with several other users.

Although the transparency of transaction information causes a series of problems, e.g., malicious uses of information, as presented in \cite{abramova2017mixing} and \cite{malinova2017market}, it enables the entrant, i.e., the new blockchain user, to possess an endogenous high reputation, i.e., the ability of performing the reliable trading. Thereby, potential users have more incentive to enter the trading system compared with the traditional real-world trading system where only the user that has high reputation can attract customers to trade with. Although the blockchain-based trading system facilitates the entry for potential entrant, the trading competition in the system increases and collusion among the users arises due to the information transparency. Thus, each potential entrant chooses whether or not to enter the system regarding the trade-off between the expected utility after the entry and the enhanced competition as well as collusion, given the other entrants' strategies. Since the potential entrant makes its choice repeatedly in discrete time period, a repeated game can be applied as presented in \cite{cong2018blockchain}. The entrant's utility is a function of the probability that one customer joins the trading in a time period, the probabilistic distribution of the reputation, the profit that can be obtained by the trading and the cost of entry. Since both the utility of the entrant and the social welfare of the system are higher than those of the traditional trading system, each potential entrant entering the system is proved to be the Nash equilibrium. However, the balance between transparency and privacy of the blockchain trading system still remains a topic for further research.

\subsubsection{Crypto-currency Value}

\begin{table*}[!htbp]
\caption{\label{tab: }A Summary of Game Theoretical Applications for Crypto-currency Economic.}
\label{Economic_summary_table}
\begin{centering}
\begin{tabular}{|>{\centering\arraybackslash}m{0.3cm}|>{\centering}m{0.5cm}|>{\centering\arraybackslash}m{1.9cm}|>{\centering}m{2cm}|>{\centering}m{2.4cm}|>{\centering}m{3.5cm}|>{\centering}m{2.2cm}|>{\centering}m{2cm}|}
\hline
 \cellcolor{mygray}  \textbf{\noun{}} &  \cellcolor{mygray} \textbf{\noun{Ref.}}&  \cellcolor{mygray} \textbf{\noun{Game Model}}&  \cellcolor{mygray} \textbf{\noun{Player}}&  \cellcolor{mygray} \textbf{\noun{Action}}  &  \cellcolor{mygray} \textbf{\noun{Strategy}} &  \cellcolor{mygray}\textbf{\noun{Payoff}} & \cellcolor{mygray} \textbf{\noun{Solution}} \tabularnewline
\hline
\hline

\parbox[t]{2mm}{\multirow{2}{*}{\rotatebox[origin=c]{90}{ \hspace{-5 cm} Crypto-currency Economic}}}
&\cite{abramova2017mixing}& Sequencial game & Blockchain users & Setting transaction transparency & Selection of mixing payments & Profits minus cost & Subgame perfect Nash equilibrium \tabularnewline \cline{2-8}
&\cite{malinova2017market}& Repeated game & Blockchain users & Setting transaction transparency & Selection of performing trading or not & Function of expected profits and possible loss & Nash equilibrium  \tabularnewline \cline{2-8}
&\cite{cong2018blockchain}& Repeated game & Blockchain users & Setting transaction transparency & Chosen of entering the system or not & Profits minus cost & Nash equilibrium \tabularnewline \cline{2-8}


&\cite{spiegelman2018game}& Potential game & Miners & Determination of the crypto-currency value & Selection between keeping mining and switching to mine on another coin & Crypto-currency value and mining rewards & Nash equilibrium \tabularnewline \cline{2-8}
&\cite{cong2018tokenomics}& Extensive-form game & Blockchain users & Determination of the crypto-currency value & Chosen of entering the system or not & Profits minus cost & Markov equilibrium \tabularnewline \cline{2-8}
&\cite{pagnotta2018equilibrium}& Non-cooperative game & Blockchain users & Determination of the crypto-currency value & Determination of the allocation of real money and investment in computational power & Function of computational power and population size of users & Nash equilibrium\tabularnewline \cline{2-8}

\hline

\end{tabular}
\par\end{centering}
\end{table*}

In the last decade, hundreds of crypto-currencies are adopted in the worldwide financial market. Each crypto-currency has its value which depends on its transaction rate, transaction fees, mining rewards and its fiat exchange rate. The miners need to choose a certain currency to mine according to the value of the crypto-currency and the competition from the other miners. Given the other miners' strategies, the miner can choose to keep mining on the same crypto-currency or change its strategy to mine on another one. Since the incentive of all miners, i.e., players, to change their strategy can be expressed using a single global function, i.e., the potential function \cite{monderer1996potential}, the potential game can be applied as presented in \cite{spiegelman2018game}. The potential function is of the distribution of the miners on mining different crypto-currencies, the computational power, the value and the reward allocation of the crypto-currencies. By using the induction of the better-response learning algorithm \cite{monderer1996potential}, the game is proved to admit more than one Nash equilibrium. However, how to achieve a desired equilibrium is not discussed.

The authors in \cite{cong2018tokenomics} further investigate the relationship between the value of the crypto-currencies and the population size of the users. Given a certain blockchain-based crypto-currency, the user can choose whether or not to participate in the blockchain platform with a cost and to hold a certain amount of the crypto-currency, given the other users' strategies. Since the user makes its strategy based on the productivity of the blockchain platform, i.e., the state which represents the quality or the usefulness of the blockchain platform, an extensive-form game can be adopted to analyze the interaction among the users as presented in \cite{cong2018tokenomics}. The user's utility is a function of the transaction supply and demand, the size of the blockchain users, the participation cost and the profit of holding the crypto-currency. By exploiting the characteristics of the Hamilton-Jacobi-Bellman (HJB) equation \cite{bardi2008optimal} transformed from the user's utility, the game is proved to admit a unique Markov equilibrium. At the equilibrium, the high crypto-currency value attracts more potential users to participate which reflects the future growth of the user population size, and the expectation of future growth leads reciprocally to a higher crypto-currency value.

Similar to \cite{cong2018tokenomics}, the authors in \cite{pagnotta2018equilibrium} demonstrate that the value of the crypto-currency is derived by the computational power of the blockchain network and the population size of the users. The user determines the amount of real money to be allocated in the transaction in the blockchain, and the miner determines the investment in the computational power in exchange for the mining profit according to the strategies of both the other users and miners. A non-cooperative game can thus be applied as presented in  \cite{pagnotta2018equilibrium}. The larger number of users attract more investment in computational power, and more computational power means the stronger consensus within the blockchain network and the higher crypto-currency value, and thereby leads to more users to participate in the blockchain network. Thus, the reciprocal interaction between the computational power and the user population size captures the equilibrium value of the crypto-currency. This equilibrium value of crypto-currency depends on the users' preferences, e.g., the risk aversion and the censorship aversion, and the usefulness of the network. The empirical data from Blockchain.info \cite{blockchaininfo} supports the theoretical analysis.

\subsection{Energy Trading}

Increasing distributed renewable energy users, e.g., solar rooftops and energy storage units, gradually changes the centralized structure of conventional power system. The reason is that the distributed energy users produce the energy and thereby users can trade their energy with each other directly. Therefore, by utilizing the decentralized structure of the blockchain network for trading information exchange, the blockchain-based energy trading systems are proposed. Each energy user in the system can decide the amount of energy to (i) buy from the conventional power system, (ii) buy renewable energy from other users, (iii) store its harvest energy, and (iv) sell its energy to the other users.

When the energy exchange price is set by the users, the interactions among the users can be modeled as games. For example, in \cite{ghosh2018exchange}, a potential game \cite{monderer1996potential} is applied to achieve the social optimum. Considering the energy demand variation, a non-cooperative game is adopted in \cite{noor2018energy}. The authors in \cite{li2017consortium} propose a credit-based energy trading system and model the interaction between the users and the credit bank as a Stackelberg game. Otherwise, when the energy exchange price is set by the system where the users bid for the exchange price, the auction models can be applied to achieve the social optimum as presented in \cite{mengelkamp2018blockchain} and \cite{kang2017enabling}.

\section{Challenges and Future Directions}
\label{sec_future_direction}
In Sections IV, V, and VI, we provide an in-depth survey on applications of game theory to address a wide range of issues in the blockchain networks and related systems. However, with the fast evolution of the blockchain technologies and their applications, a plethora of emerging problems remain open for further studies, many of which can be solved using game theory. In this section, we expand our discussion to some challenges as well as research directions with blockchain, where the mathematical tools of game theory may exert further potential for system analysis and mechanism design.

\subsection{Challenges from Game Theory Perspective}
\subsubsection{Existence of Nash Equilibria}

Most references reviewed in this survey discuss the existence of the unique Nash equilibrium. At the Nash equilibrium, the players, e.g., the miners or the pools, have no
incentive to deviate from their current strategies. However, in practice, multiple Nash equilibra can exist, and thus it is challenging for the players to choose the optimal strategy or solution. For example, for the mining management \cite{easley2017mining}, with the existence of Nash equilibra, the miners can choose between staying and leaving, and the blockchain users choose between paying or not paying the transaction fee. In this case, finding the solution among the Nash equilibria to achieve a social optimum for the whole network is a challenge. Similarly, for the crypto-currency economic \cite{spiegelman2018game}, how to achieve a social optimal equilibrium in the crypto-currency market is very challenging.

\subsubsection{Implementation of Game Models}
The applied game models proposed in aforementioned reviews have its limitation. For example, due to the first-mover advantage, the Stackelberg game is widely used to solve many issues in blockchain network. However, the blockchain network is a type of decentralized system with a number of distributed nodes, i.e., players. Therefore, how leader nodes observe the strategy of each follower node, make optimal decisions, and find the equilibrium is one big challenge. To address the challenge, the meanfield games \cite{lasry2007mean} can be applied for analyzing the performance of the whole blockchain network with large number of miners where individual miners have relatively negligible impact upon the network. In addition, evolutionary games can be adopted for analyzing mining pools' formation and evolution. Stochastic games can be used for analyzing more complex scenarios, such as miners' probabilistic selection of transactions to be included, blocks to be verified and broadcast, and chains to be attached and mine.

\subsection{Open Issues and Research Directions for Applications of Game Theory in Blockchain}
\subsubsection{Throughput Improvement}
Blockchain technologies have been adopted in a number of scenarios. However, the throughput, i.e., capacity of processing requested transactions, of blockchain networks limits the scope of blockchain applications. The major reasons for this issue are the long block creation time and limited block size \cite{houy2014bitcoin}. However, block creation time and the block size cannot be easily changed for improving the throughput. The analyses in \cite{zhang2017necessity} show that miners intend to form a coalition if the block size is unlimited. This is harmful to maintaining the decentralized structure of the blockchain network. Also, the authors in \cite{easley2017mining} demonstrate that even with the unlimited block size, there is still a limitation on throughput imposed by the waiting time for transactions to be included in blocks. Thus, to improve the throughput, blockchain protocols in terms of the efficient block creation and the proper block size need to be further developed, and game theory can be a useful tool.

\subsubsection{Alternative Consensus Mechanisms}
In blockchain networks, e.g., PoW networks, every node performs several certain tasks to maintain the consensus across the blockchain. However, reaching the consensus needs nodes to repeat tasks consuming a large amount of electricity \cite{dimitri2017bitcoin}. Thus, an alternative consensus mechanism to PoW such as Proof of Useful Work or Resources (PoUWR) \cite{bottou1991stochastic} may be used. For example, computing hash value in PoW network can be replaced with performing stochastic gradient descent for neural network training \cite{bottou1991stochastic}. Due to the difference between the tasks in terms of data volume, expected accuracy and variable dimension, the strategies of nodes to obtain a puzzle solution are different from those in the PoW network. Therefore, it is necessary to apply game approaches to analyze the interaction among nodes in the process of PoUWR competition, e.g., the computational power allocation between PoUWR and PoW, the tradeoff between the payoff and the cost, and security issues regarding the deviation from the PoUWR protocol.

\subsubsection{Permissioned Ledger Types}
Public blockchain has been adopted in many applications. Public blockchain allows anyone to participate to be a node, and it has no control by regulatory agencies, industries, or governments. In addition to the public blockchain, permissioned blockchain ledger types such as consortium blockchain, become another interesting application    of Nakamoto's blockchain implementation. Consortium blockchains can be considered to be semi-decentralization. The reason is that not everyone can participate in the network, and the consortium blockchain is maintained by a group of pre-selected nodes, allowing for a greater degree of control over the network by regulators. As such, the consortium blockchains involve multiple entities and stake-holders, i.e., the pre-selected nodes, the verification nodes, and the blockchain users. To model and analyze complex interactions among the entities and stake-holders, game theory can be adopted as a useful tool. For example, the non-cooperative games can be used to analyze the pre-selected node selection, Stakeleberg games can be applied to analyze the interaction between the pre-selected nodes, i.e., the leaders, and the verification nodes, i.e., the followers. Also, evolutionary games can be used to analyze the formation of mining pools in permissioned blockchain networks.

\subsubsection{Incorporating Blockchain Technologies into Other Scenarios}

As blockchain is a versatile technology, it is also possible to incorporate blockchain into other emerging network and application scenarios. For example, the authors in \cite{xiong2018optimal} introduce a blockchain-based edge computing paradigm in which mobile users offload their computing tasks to computing service providers and pay the corresponding fees. This paradigm addresses the implementation issue of blockchain applications on resource-limited mobile services. However, the blockchain-based edge computing paradigm raises resource management issues. For example, how to motivate the service providers to contribute their computing resources. Game theory can be efficiently used to design incentive mechanisms. For example, auction schemes can be adopted to improve the utility or revenue of the service providers. Also, the Stackelberg game can be applied to improve both the utility of the computing service providers and the mobile users. Predictably, by taking advantage of game theory to analyze and design incentive mechanisms, blockchain technologies can be widely incorporated into multi-agent scenarios beyond the crypto-currencies, e.g., mobile blockchain networks, information sharing scenarios, and energy trading markets.

\section{Conclusions}

This paper has presented a comprehensive survey of the applications of game theory in blockchain. Firstly, we have given an overview of blockchain with its structure, workflow, and incentive compatibility. Then, we have introduced the basic knowledge of game theory and several game models with the objective to understand the motivations of using game theory to analyze interactions among different components in blockchain. Afterwards, we have provided reviews and analyses using game theory in detail to deal with a variety of problems regarding security, mining management and blockchain applications. Finally, we have outlined existing challenges as well as future research directions.

\ifCLASSOPTIONcaptionsoff
  \newpage
\fi



%
%
%
\bibliographystyle{IEEEtran}
\bibliography{Blockchain_Survey_Final_Draft}
%







\end{document}